\documentclass{article}
\usepackage{chicago}
\usepackage{latexsym}
\usepackage{url}
\usepackage{graphicx}

\newcommand{\iec}{\mbox{i.\,e.\,}}

\newcommand{\eg}{\mbox{e.\,g.\,\ }}

\newcommand{\opm}[1]{\ensuremath{#1}}
\newcommand{\opmad}[1]{\ensuremath{#1^\dagger}}
\newcommand{\EU}{\mathrm{EU}}

\newcommand{\mc}[1]{\ensuremath{\mathcal{#1}}}

\newcommand{\ket}[1]{\ensuremath{\left|  #1 \right\rangle}}
\newcommand{\bra}[1]{\ensuremath{\left\langle #1 \right|}}

\newcommand{\proj}[2]{\ensuremath{\ket{#1} \bra{#2}}}
\newcommand{\tpk}[2]{\ensuremath{\ket{#1}\!\otimes\!\ket{#2}}}

\newcommand{\matel}[3]{\ensuremath{\bra{#1} #2 \ket{#3}}}
 
\newcommand{\op}[1]{\ensuremath{\widehat{\textsf{\ensuremath{#1}}}}}

\newcommand{\id}{\op{\mathsf{1}}}

\newcommand{\be}{\begin{equation}}
\newcommand{\ee}{\end{equation}}

\newcommand{\Proof}{\noindent \textbf{Proof: }}
\begin{document}

\title{A formal proof of the Born rule from decision-theoretic assumptions}
\author{David Wallace}
\maketitle
\begin{abstract}I develop the decision-theoretic approach to quantum probability, originally proposed by David Deutsch, into a mathematically rigorous proof of the Born rule in (Everett-interpreted) quantum mechanics. I sketch the argument informally, then prove it formally, and lastly consider a number of proposed ``counter-examples'' to show exactly which premises of the argument they violate.

(This is a preliminary version of a chapter to appear --- under the title ``How to prove the Born Rule'' --- in Saunders, Barrett, Kent and Wallace, \emph{Many worlds?   Everett, quantum theory and reality}, forthcoming from Oxford University Press.) 
\end{abstract}

\section{Introduction}

\begin{quote}
Thus we see that quantum theory permits what philosophy would hitherto have regarded as
a formal impossibility, akin to ``deriving an ought from an is'', namely deriving a
probability statement from a factual statement. This could be called deriving a ``tends
to'' from a ``does''. \cite{deutschprob}
\end{quote}

The ``Everett interpretation of quantum mechanics'' is just unitary quantum mechanics, taken literally as a description of the world; it is a ``many-worlds'' theory because it instantiates multiple, emergent, branching quasiclassical realities. That much is commonplace amongst contemporary Everettians; it was argued for \emph{in extenso} in my other chapter in this volume, and for the purposes of \emph{this} chapter I will take it as read.

It is widely held, however, that a problem remains: namely, how does \emph{probability} fit into this story? It is not in dispute what physical magnitude is \emph{supposed} to be (or to stand in for) probability: the probability of a branch is supposed to be its weight (\iec, its mod-squared amplitude). More formally, the probability of a history $\alpha$ represented by a history operator $\op{C}_\alpha$ (in the consistent-histories formalism)
is supposed to be
\be
\Pr(\alpha)=\matel{\psi}{\op{C}_\alpha}{\psi}
\ee
where \ket{\psi} is the universal (Heisenberg-picture) state. In more parochial language, if an observer's branch has weight $w_0$, if it is going to split into multiple branches, and if those branches in which $X$ happens have weight $w_X$, then the probability for that observer of $X$ happening is supposed to be $w_X/w_0$.

What is in dispute is why, and how, this physical magnitude can be probability. One might ask: how can it even \emph{make sense} for anything to ``be probability'' in a theory where all possible outcomes occur; less confrontationally, one might ask what kind of \emph{argument} can be given to justify the claim that mod-squared amplitude is probability. (I have previously referred to these as the \emph{Incoherence Problem} and the \emph{Quantitative Problem} respectively, though I shall not make much of the distinction in this article.)

One quite legitimate response to this question, I think, is bemusement. After all, formally speaking the measure defined by mod-squared amplitude on any given space of consistent histories 
satisfies the algorithms for a probability. Indeed, mathematically the setup is identical to any stochastic physical theory, which ultimately is specified by a measure on a space of kinematically possible histories (albeit that measure is usually given indirectly, via a stochastic differential equation). In Everett-interpreted quantum mechanics, the correct space of consistent histories is the space of quasi-classical histories; as my earlier chapter argued, this space may be imprecisely defined but this is no reason not to take it seriously as emergent structure. So (goes the response) all the formal requirements to take mod-squared amplitude as probability are in place; to ask for more is no more justified than to ask why the physical quantity represented by the metric in Newtonian space ``is'' length, or why other mathematical features of classical physics represent mass or charge. (\citeN{saundersprobability} develops this response in more depth.)

I have a good deal of sympathy for this response, but in this chapter I wish to discuss a more positive answer to the question, which might be called the \emph{decision-theoretic strategy}.
\begin{description}
\item[Decision-theoretic strategy:] Probability gets its meaning in quantum mechanics through the rational preferences of agents. In particular, a rational agent who knows that the Born-rule weight of an outcome is $p$ is rationally compelled to act as if that outcome had probability $p$.
\end{description}

The decision-theoretic strategy was advocated by \citeN{deutschprob}. In the same paper he presented an informal proof, 
from principles of decision theory that (he argued) did not themselves invoke probabilistic notions, that the only rational strategy for an agent in an Everettian universe is to follow the Born rule. I developed Deutsch's proof further, and presented alternative versions of it, in \citeN{decshort} and \citeN{wallaceprobdec}. 

The argument has met with its share of criticism. Some of the criticism (\eg Barnum~\emph{et al}~\citeyear{barnumetal}, \citeN{lewisondeutsch}) has been directed at the proof itself; some (\eg Albert and Price's contributions to this volume) has sought to undermine the possibility of a proof by proposing other, (allegedly) equally rationally justifiable alternatives to the Born rule. (These, I should make clear, are not proposed as positive \emph{suggestions} as to how we should act if the Everett interpretation is correct; they are intended rather as \emph{reductios}.)

My purpose in this chapter is to give a self-contained defence of the decision-theoretic strategy, culminating in a formal proof of the rational necessity of the Born rule from axioms of decision theory which I will defend. Formalisation is not often an aid to understanding, but when a result is controversial it can be helpful to see \emph{exactly} what is and is not required. Along the way I will showcase some of the various proposed alternative strategies for Everettian rationality, and show exactly how they conflict with the assumptions in the argument; in doing so, I hope, the argument for making those assumptions will become clearer.

My focus in the chapter is deliberately narrow. A proof that rational agents in an Everett universe must act in accordance with the Born-rule probabilities falls short of a full solution of the probability problem: we might also ask how this decision-theoretic notion of probability connects with our use of probability in assessing the evidence for quantum mechanics, or with our ordinary, pre-theoretic notions of probability as a guide to action in cases of uncertainty. I shall address neither question here, though (for discussions of the former question, see Greaves and Myrvold's contribution to this volume, \citeN{wallaceepist}, and part II of \citeN{evbook}; for the latter, see \citeN{wallacebranching}, part III of \citeN{evbook}, \citeN{saundersprobability}, and Saunders' contribution to this volume.)

I shall begin in section~\ref{wallace-decpreamble} with a brief discussion of decision theory \emph{in general}, and go on in section~\ref{wallace-deceverett} to see how decision theory works in the Everett interpretation. The main part of the paper (sections~\ref{wallace-informalstatement1}--\ref{wallace-formalproof}) states and proves, first informally and then in full mathematical rigor, that the Born rule is the unique rational strategy available in the Everett interpretation. I illustrate this result (section~\ref{wallace-otherstrategies}) by considering a number of other proposed strategies, proposed at various times and places as counterexamples to the necessity of the Born rule, and show why those strategies are not in fact valid alternatives. I conclude with a few more general observations about why the Born rule can indeed be proven, and why the Everett interpretation is essential in such a proof.

\section{Preamble: the decision-theoretic approach}\label{wallace-decpreamble}

Suppose an coin is to be tossed in five minutes' time; and suppose that an agent bets five dollars (at even odds) that it will land heads. There are two interestingly different possible results: (i) the measurement gives result ``up'' and the agent gets $\$5$; (ii) the measurement gives result ``down'' and the agent loses $\$5$. If the result is heads, the agent will be pleased about the bet; if it is `tails he will be less delighted, though (if he is of an appropriate character) he may well still regard the bet as having been the right choice given his information before the coin toss. (There are of course vastly more than two microscopically distinct possible results; the division into two sets is based on the pragmatic interests of the agent.)

In deciding whether to accept this bet as opposed to any number of other bets, the agent has to weigh the cost to him if the result is ``down'' against the benefit if it is ``up''. Decision theory gives a precise answer to the question of how he should carry out this weighting: he should assign some utility (some real number) $\mc{V}(+\$5)$ to receiving five dollars, and some other utility $\mc{V}(-\$5)$ to losing five dollars, and some third utility $\mc{V}(\$0)$ to neither getting nor losing it; and he should assign a probability $\Pr(H)$ to heads; and he should take the bet only if
\be
\Pr(H)\times \mc{V}(+\$5)+(1-\Pr(H)) \times \mc{V}(-\$5)>\mc{V}(\$0).
\ee
More generally, decision theory mandates that an agent should assign a utility to each payoff, and a probability to each outcome, and that faced with \emph{any} decision, the agent should choose that option which maximises expected utility with respect to those assignments.
 
(In elementary discussions, it is common just to assume $\mc{V}(+\$N)=N$, but this is too simplistic: it is not irrational to refuse to trade your house for a one-in-a-thousand chance of winning Microsoft. In fact, in a decision-theoretic framework, what it \emph{means} to say that one reward is twice as valuable as another is that a $50\%$ chance of getting the first is as valuable as getting the second with certainty. See \citeN[pp.91--104]{savage} for more on this point.)

\emph{Why} should an agent behave this way? Prima facie, it isn't obvious at all that he should try to maximise expected utility rather than, say, maximising utility with respect to the square of the probability function; or maximising the logarithm of utility; or just maximising the utility of the least good possible outcome.

Decision theory has a standard answer to this question: if an agent has a definite preference (which might be indifference) between any two bets, and if that preference order obeys certain constraints which are purported to be necessary conditions of rationality, and if the set of available bets has a sufficiently rich structure, then it is possible to prove a \emph{representation theorem}: a theorem that for any such preference order there is a unique probability function, and an essentially unique\footnote{By ``essentially unique'' I mean unique up to positive affine transforms $x \longrightarrow ax+b$, with $a$ positive. Fairly obviously, such transformations serve only to scale the expected utility of all bets in the same way.} utility function, such that one bet is preferred to another iff it has a higher expected utility. It follows that any agent whose preferences cannot so be represented must be acting irrationally: that is, must somewhere be violating a principle which (again, purportedly) is a necessary constraint on rational action.

It will be instructive to present two such principles (both drawn from \citeN{savage}). The first is \emph{transitivity}: if an agent prefers $a$ to $b$ and $b$ to $c$, he should prefer $a$ to $c$. The second might be called \emph{dominance}: if an agent will do better through $a$ than $b$ \emph{whatever happens}, he should chose $a$ over $b$. (For instance, a bet that pays ten dollars if a coin-toss lands heads and nothing if it lands tails is to be preferred over a bet on the same toss that pays five dollars on heads and minus five on tails).

There is an important weakness in this decision-theoretic argument which needs to be stressed. It is a proof  that rational agents must bet according to \emph{some} probability function, but it is silent on the connection between that function and the ``real'' probabilities. No decision-theoretic principle is contradicted by an agent who assigns probability $99/100$ to an apparently-fair coin landing ``heads'', for instance. A minority of advocates of the decision-theoretic approach simply deny that there is any such thing as objective or ``real'' probability; the majority just take it as a bare postulate that an agent should conform his subjective probabilities to the objective probabilities when he knows the latter.\footnote{This is a simplification: a more general statement is that an agent's subjective probability in $X$ conditional on the objective probability of $X$ being $p$ should in turn be $p$ (this is known as the Principal Principle, following \citeN{lewischance}). My main point stands, however: the Principal Principle is postulated, not derived.}

This weakness actually rather undermines the use of probability as a criticism of the Everett interpretation, even without the arguments of this paper: if classical probability can give no justification of its probability rule, why ask the Everettian for such a justification? But in fact, we will see that in the Everett interpretation, not only can we use rationality considerations to make sense of probabilities as well as in conventional decision theory, we can prove and not merely postulate the link between those probabilities and the quantum-mechanical weights.

\section{Everettian Rationality}\label{wallace-deceverett}

So: consider the Everettian version of the coin-toss. Instead of a coin, we have a particle in a superposition of spin-up and spin-down (in some fixed direction); instead of a coin-toss, we have a spin measurement. And instead of there being two interestingly different \emph{possibilities}, there are two interestingly different sets of branches: the spin up branches, where (if the agent took the bet) he gets five dollars, and the spin down ones, where he loses five dollars. In deciding whether to accept this bet as opposed to any number of other bets, the agent has to weigh the benefit to himself in the branches where the result is up against the cost in the branches where the result is down. So the notion of a bet is at least meaningful in the Everettian context.
\begin{description}
\item[Skeptic:] The benefit isn't to the agent: it's to copies of that agent in the future.
\item[Author:] Sure. But that's true in the non-Everettian just as in the Everettian case. In either case, the reason the agent makes one choice rather than another is because of his concern about his future interests --- that is, about the interests of his future self or selves.

And an agent's future self \emph{is} his future self just by virtue of the causal, structural, dynamical relations between it and the agent's past self. There is (I assume!) no indivisible, immaterial soul which passes through my life and magically makes me a single being: what makes the stages of me at different times all \emph{me} is that they are appropriately related. And it seems, at least, that an Everettian agent's future selves stand in all the same relations to him as a non-Everettian agent's future selves stand to \emph{him}. 
\item[Skeptic:] There's a pretty obvious disanalogy, though. In the Everettian case, there's more than one future self!
\item[Author:] Fair enough. (Though there are some subtleties here: see \citeN{saunderswallace}, and also Saunders' chapter in this volume, for a construal of personal identity in which this is not the case.) But it's hard to see why, in the Everett case, I should regard my future self as any the less \emph{me} --- why I should not treat his goals and desires, his hopes and dreams, as my own --- just because I actually have multiple such selves. And so it's hard to see why those future selves should be less relevant to my considerations now --- to my decision-theoretic preferences --- than would be the case in the absence of Everettian branching.
\end{description}

So an Everettian agent can be in a decision problem --- can be faced with a choice of bets --- just as can a non-Everettian. And certainly \emph{one} strategy available to him is what might be called the ``Born-rule strategy'': choose that bet which maximises expected utility with respect to the Born-rule weights (that is, the mod-squared amplitudes). An Everettian agent who adopted the Born-rule strategy would make exactly the same choices between bets as would a non-Everettian who adopted the Principal Principle with respect to the Born-rule weights. The two would be indistinguishable in terms of their behavioural dispositions. 

Is it the \emph{only} strategy? Advocates of Deutsch's decision-theoretic strategy say that it is. More precisely, they argue that given certain principles of rationality, and given knowledge of quantum mechanics, it can be proved that any strategy other than the Born-rule strategy violates some rational constraint on action. By analogy with the Representation Theorems of classical decision theory, we might call such a result a (purported) \emph{Quantum Representation Theorem}. 

Such a theorem can in fact be proved formally, and in sections~\ref{wallace-formalstatement}-\ref{wallace-formalproof} I will give a formal proof of such a theorem. But since a fully formalised proof of a result does not make for accessible reading, firstly I will give an \emph{informal} version. In sections~\ref{wallace-informalstatement1}-\ref{wallace-informalstatement2} I will state informally, and motivate, the axioms I wish to use; in section \ref{wallace-formalstatement} I will argue informally why these axioms jointly entail a Quantum Representation Theorem. 

\section{The quantum decision problem}\label{wallace-informalstatement1}

The situation I wish to consider is the following. A quantum state is to be prepared in some superposition; the system is measured in some basis; a bet is made by the agent on the outcome of that measurement. Our agent knows (we assume) that the Everett interpretation is correct; he is also assumed to know the universal quantum state, or at least the state of his branch. (The latter is an unrealistic but convenient assumption; in practice, however, it suffices for the agent to know the mod-squared amplitudes for each outcome of a measurement.) His preferences can be represented by an ordering relation on these bets.

Since (in Everettian quantum mechanics, at any rate) preparations, measurements, and payments made to agents are all physical processes, there is a certain simplification available: any preparation-followed-by-measurement-followed-by-payments can be represented by a single unitary transformation. So our agent's rational preference is actually representable by an ordering on unitary transformations. 

We should acknowledge that not all unitary transformations represent something physically possible.\footnote{Doesn't \emph{only one} unitary transformation represent something physically possible? Doesn't the Hamiltonian of the universe uniquely determine which transformation is performed? If this is a problem, it is not specific to Everett: it is the ancient debate of free will vs.\,determinism. Rather than get into this morass (though I recommend \citeN{dennettelbowroom} for reassurance that the two are compatible), let me just note that we can talk about \emph{rational strategies} even if an individual agent is not free to choose whether or not his strategy is rational.} In particular, transformations which lead to \emph{recoherence} --- that is, to Everett branches merging --- are certainly not performable by any agent localised to a specific branch. But nonetheless we will consider a fairly wide set of transformations to be available --- exactly how wide is something that the axioms will spell out. (It might be worth recalling at this stage that decision theory is concerned with the preferences an agent would have when confronted with a particular decision --- his dispositional preferences, in philosophers' language --- and not just with what actually happens. It is most unlikely that I will be offered a choice between the presidency of the World Bank and the deputy leadership of the Al-Qaeda terror group, but I have a definite preference between the two. As such, the assumption of a reasonably wide set of transformations seems reasonable enough.)

We should also acknowledge in our decision-theoretic setup that decoherence imposes a certain structure on the Hilbert space. We can represent this by a resolution of the identity on the Hilbert space: that is, by a decomposition of the space into subspaces, with each subspace $\pi$ corresponding to a possible macrostate. The choice of macrostates is largely fixed by decoherence, although the precise fineness of the grain of the decomposition is underspecified. (In the model, of course, it will be precisely specified, but this just illustrates that the model is artificially precise.) We call a macrostate \emph{available} to an agent if there is an available act which, when performed, leaves some of his future selves in that macrostate. 

Part of the point of the decomposition into macrostates is that an agent can be assumed not to care exactly what the microstate is within a given macrostate (if he does care, we have defined the macrostates too coarsely). But in fact, usually an agent will also be indifferent between a great many macrostates: for instance, if offered a million dollars, I am indifferent as to the colour of the cheque.\footnote{The reader who doubts this claim is encouraged to test it empirically.} It will be useful to consider a coarse-graining of the macrostate subspaces into \emph{reward subspaces}, such that an agent's only preference is to which reward subspace he is in. Formally speaking, ``reward subspace'' is a derived concept within the decision theory. 

In fact, for mathematical reasons it will be convenient to work both with the set of macrostates and with the Boolean algebra \mc{E} of arbitrary disjunctions of macrostates\footnote{Recall that the disjunction $E\vee F$ of two subspaces of a Hilbert space is the closure of the span of their union.}, which we call the \emph{event space}. The formal development of the theory will not actually require the assumption that the event space can be constructed from a set of macrostates (though it does not rule out that assumption). Indeed, since the fineness of grain of branches is indeed underspecified, the branch structure might be best idealised in some particular situation by a model in which the algebra is not constructed this way. (For instance, if the Hilbert space is $L^2(R^N)\otimes \mc{H}_E$, where $\mc{H}_E$ represents some subsystem of environmental degrees of freedom, then we might wish to take the elements of \mc{E} to be the subspaces
\be 
\Sigma_E=\{f\otimes v:\,E\mbox{ is an open subspace of }R^N\mbox{ and } f \mbox{ has support in E}\}
\ee  
which cannot be generated from macrostates (unless we are willing to relax rigor and consider eigenstates of position).

For simplicity, we will refer to the set of unitary transformations over which an agent's preference order is defined as \emph{acts}. A different set may be relevant for different physical states of the universe, so we will have cause to speak of the \emph{acts available at } a macrostate $\pi$. (In view of the previous paragraph's comment, we might do better to talk of the acts which are \emph{contemplatable} at $\pi$; I avoid this terminology mostly because it's cumbersome.) 

In fact, it will be simpler to talk of which acts are available at a given event (not just a macrostate) --- informally an act available at an event $E=\pi_1\vee \pi_2 \vee \cdots \vee \pi_N$ is the conditional act `if the macrostate is actually $\pi_i$, perform $\opm{U}_i$. This makes it much more straightforward to talk about the composition of acts: if \opm{U} is available at an event $E$, and \opm{V} is available at the smallest event containing the range of $\opm{V}$, for instance, then $\opm{V}\opm{U}$ ought to be the act of performing \opm{U} and \opm{V} sequentially and so also should be available at $E$. In the formal development we will state explicit rules to ensure that these and similar compositions are available; for now we take it as tacit that they are.

We now need to represent the agent's preferences between acts. Since those preferences may well depend on the state, we write it as follows:
if the agent prefers (at $\psi$) act $\opm{U}$ to act $\opm{U}'$, we write
\be
\op{U}\succ^\psi \op{U}'.
\ee
To be meaningful, of course, this requires that $\opm{U}$ and $\opm{U}'$ are both available at $\psi$'s macrostate. So $\succ^\psi$ is to be a two-place relation on the set of acts available at that macrostate. In the event formalism we use later, we will require $\succ^\psi$ to be a two place relation on the acts available at each event which contains $\psi$.

So much for the setup; now for the axioms. They come in two categories: axioms of \emph{richness}, which concern which acts are available to the agent (how rich the structure of the set of acts is) and which are not connected to a particular agent's preference order; and axioms of \emph{rationality}, which constrain that preference order. 

The richness axioms, then, are
\begin{description}
\item[Reward availability:] All rewards are available to the agent at any macrostate: that is, the set of available acts always includes ones which give all of the agent's future selves the reward. 
\item[Branching availability:] Given any set of positive real numbers $p_1, \ldots p_n$ summing to unity, an agent can always choose some act which has $n$ different macrostates as possible outcomes, and gives weight $p_i$ to the $i$th outcome. 
\item[Erasure:] Given a pair of states $\psi\in E$ and $\varphi\in F$ in the same reward, there is an act $\op{U}$ available at $E$ and an act $\op{V}$ available at $F$ such that $\op{U}\psi=\op{V}\varphi$.
\item[Problem continuity:] For each event $E$, the set of acts available at $E$ is an open subset of the set of unitary transformations from $E$ to \mc{H}.
\end{description}
These should mostly be uncontroversial. Branching availability and reward availability are consequences of the relatively stylised decision problem we are considering, where measurements are being made and payments are being provided; they reflect the facts (respectively) that quantum systems can be prepared in arbitrary states and that envelopes of cash can always be given to people.

Erasure is slightly more complicated. It effectively guarantees that an agent can just forget any facts about his situation that don't concern things he cares about (that, is, by definition: that don't concern where in the reward space he is). In thinking about it, it helps to assume that any reward space has an ``erasure subspace'' available (whose states correspond to the agent throwing the preparation system away after receiving the payoff but without recording the actual result of the measurement, say). An ``erasure act'' is then an act which takes the quantum state of the agent's branch into the erasure subspace; the agent is (by construction) indifferent to performing any erasure act, and since he lacks the fine control to know which act he is performing, all erasures should be counted as available if any are. It follows that, since for any two such agents all erasures are available, in particular there will be two erasures available satisfying the axiom. 

I postpone a discussion of problem continuity until the axioms of rationality have been introduced.

\section{The dictates of rationality}\label{wallace-informalstatement2}

Moving on to the rationality axioms, they come in two groups. The first two axioms are very general principles of rationality, as relevant in the classical as in the quantum context.
\begin{description}
\item[Ordering:] The relation $\succeq^\psi$ is a total ordering for each $\psi$ on the set of acts available at $\psi$, for each $\psi$ (that is: it is transitive, irreflexive and asymmetric, and if we define $\opm{U}\sim^\psi \opm{V}$ as holding whenever $\opm{U}\succ^\psi \opm{V}$ and $\opm{V}\succ^\psi \opm{U}$ fail to hold, then $\sim^\psi$ is an equivalence relation).
\item[Diachronic consistency:] If \opm{U} is available at $\psi$, and (for each $i$) if in the $i$th branch after $\opm{U}$ is performed there are acts $\opm{V}_i,\opm{V}'_i$ available, and (again for each $i$) if the agent's future self in the $i$th branch will prefer $\opm{V}_i$ to $\opm{V}'_i$, then the agent prefers performing $\opm{U}$ followed by the $\opm{V}_i$s to performing $\opm{U}$ followed by the $\opm{V}'_i$s.
\end{description}
Ordering is utterly familiar (indeed, built in to our use of the $\succ^\psi$ symbol) and hopefully uncontroversial. But it is worth stressing that the \emph{reason} it is uncontroversial is not (just!) that it would be unintuitive for an agent's preferences to violate ordering, but because it isn't even possible, in general for an agent to formulate and act upon a coherent set of preferences violating ordering. 

Of course, in stylised and artificial special cases, it might be. If an agent knows that he will be offered three acts chosen from a set of ten, he can arbitrarily pick one element from each three-element subset, and elect to choose that one. But of course, real decision problems aren't that cleanly specified: the precise number of acts available is vague or just indeterminate and the cognitive cost of trying to pin down the size of that set is prohibitive (even when the very act of trying to pin it down does not change the problem out of recognition). Excluding stylised and occasional exceptions, then, ordering is \emph{constitutive} of rationality, not just intuitively necessary for it.

I have stressed this because, in fact, very much the same defence can be offered of the less-familiar diachronic consistency principle, which in effect rules out the possibility of a conflict of interest between an agent and his future selves. In philosophy examples one often speaks of a (classical) agent as if he were a continuum of independent entities, one for each time, each having his own preference ordering. But of course actual decision-making takes place over time. An agent's actions take time to carry out; his desires and goals take time to be realised. If his preferences do not remain consistent over this timescale, deliberative action is not possible at all.

Of course, there are plenty of \emph{localised} violations of diachronic consistency even outside the Everettian context. If I tell my friend not to let me order another glass of wine after my second, I acknowledge that my desires at that point will conflict with my desires now. But notice that such situations
\begin{description}
\item[(a)] are generally not taken to be rational;
\item[(b)] are indeed analysed as situations of conflict, where my present self acts to prevent my future self having access to his preferred choice;
\item[(c)] are localised, taking place against a general assumption of diachronic consistency in myself and others (as when I assume that my friend's future self will indeed act on her agreement not to let me order the wine, or that the morning after the night before, I'll be glad that she did).\footnote{For arguments that ascriptions of irrationality \emph{only} make sense against a presumed backdrop of rationality, see Davidson~(\citeyearNP{davidsoninterpretation,davidsonparadoxes}), \citeN[pp.\,83--116]{dennettintentional}, and \citeN{lewisradical}.}
\end{description}
Similarly: in a branching universe, to accept a conflict of interest between my pre-branch and post-branch selves is to cease to see them as the same person. If branching were an isolated occurrence, this might be possible: it is arguably callous to make a copy of myself and send him off to do a dangerous or disagreeable task --- and, crucially for the point, to take actions designed to prevent him shirking that task but it is not \emph{irrational}.\footnote{See the first part of Greg Egan's novel \emph{Permutation City} for a science-fictional exploration of the idea --- but notice that its plausibility relies on the copy's actions being causally relevant to the original, something not possible in the Everettian universe.} But \emph{Everettian} branching is ubiquitous: agents branch all the time (trillions of times per second at least, though really any count is arbitrary). In the presence of \emph{widespread, generic} violation of diachronic consistency, agency in the Everett universe is not possible at all.
\begin{description}
\item[Skeptic:] Stop there. You're trying to argue that rationality (agency, if you like) even makes sense in an Everett universe. You can't do that by saying that rationality is impossible unless such-and-such. Maybe there just isn't any coherent notion of rationality in the Everett interpretation?
\item[Author:] You misunderstand. I'm just saying that rationality requires diachronic consistency: that any rational strategy is a diachronically consistent strategy. So I'm constraining the space of rationally possible behaviours. If it turns out to be empty, of course, we're in trouble. But it won't: the Born-rule strategy is diachronically consistent and satisfies all the other axioms. All I'm doing is restricting (eventually to zero) the set of non-Born strategies. 
\item[Skeptic:] What if the Born rule is also irrational?
\item[Author:] Which is to say: what if it violates some rationally required constraint on action? Then we're sunk. But it doesn't.
\item[Skeptic:] What about ---
\item[Author:] Yes, yes, ``it's rationally required to weight each branch equally''. We'll come to that. 
\end{description}
Incidentally, the very idea of composing acts to make further acts also presupposes diachronic consistency: only if an agent can think of future decisions he will make as \emph{his decisions}, so that he can meaningfully make those decisions (for all that there is always some possibility that he will change his mind) does it make sense to consider composite acts.

The remaining rationality axioms are more specific to the Everettian context. Their precise statements get a bit more technical, so I phrase them fairly loosely here; as always, see section~\ref{wallace-formalstatement} for details and for reassurance there there isn't sleight-of-hand going on.
\begin{description}
\item[Microstate Indifference:] An agent doesn't care what the microstate is provided it's within a particular macrostate. 
\item[Branching Indifference:] An agent doesn't care about branching \emph{per se}: if a certain measurement leaves his future selves in $N$ different macrostates but doesn't change any of their rewards, he is indifferent as to whether or not the measurement is performed.
\item[State Supervenience:] An agent's preferences between acts depend only on what physical state they actually leave his branch in: that is, if $\opm{U}\psi=\opm{U}'\psi'$ and $\opm{V}\psi=\opm{V}'\psi'$, then an agent whose prefers $\opm{U}$ to $\opm{V}$ given that the initial state is $\psi$ should also prefer $\opm{U}'$ to $\opm{V}'$ given that the initial state is $\psi'$ --- $\opm{U}\succ^\psi \opm{V}$ iff $\opm{U}'\succ^{\psi'} \opm{V}'$.
\item[Solution Continuity:] If for some state $\psi$ $\op{U}\succ_\psi\op{U}'$, then sufficiently small permutations of $\op{U}$ and $\op{U}'$ will not change this.
\end{description}
Macrostate indifference is hopefully uncontroversial: it's built into the definition of macrostates, in fact. (The point being that an agent can have no practical control as to what state he gets, within a particular macrostate, on familiar statistical-mechanics and decoherence grounds.)

Solution continuity and  branching indifference --- and indeed problem continuity --- can be understood in the same way, in terms of the limitations of any physically realisable agent. Any discontinuous preference order would require an agent to make arbitrarily precise distinctions between different acts, something which is not physically possible. Any preference order which could not be extended to allow for arbitrarily small changes in the acts being considered would have the same requirement. And a preference order which is not indifferent to branching \emph{per se} would in practice be impossible to act on: branching is uncontrollable and ever-present in an Everettian universe.\footnote{The main source of branching is probably classically chaotic systems; see \citeN{zurek94} for technical details, and \citeN{wallacestatmech} for discussion.}
\begin{description}
\item[Skeptic:] Why assume \emph{a priori} that the rational strategy must be physically possible? Even if there is some strategy in an Everettian universe which counts as rational, maybe it's not physically possible to carry out that strategy.
\item[Author:] That's confused. Firstly, we already know there's at least one possible rational strategy: the Born rule. Secondly, what would it even be for a strategy to be rational, but physically impossible? By that token, the rational strategy for a trader is ``always buy shares that are going to increase in value''.

To be fair, a strategy might be literally impossible but be an idealization of a possible strategy --- after all, perfect rationality itself is an idealization. One might \emph{possibly} relax the assumption of Continuity on these grounds (and I'll make some comments on that later), though I don't really think it's justified. But no strategy can approximate caring about branch number, as we'll see.
\end{description}
The other way to understand these assumptions is as prohibitions on strategies that just exploit artefacts of our model. The branching structure --- including the well-defined number of branches associated with any act --- is derived from the set of macrostates, which is in turn derived from decoherence. But as I argued in my earlier chapter in this volume that this structure has a significant degree of arbitrariness associated with it, primarily in terms of the coarseness of the grain of the macrostates (see also James Hartle's contribution). Put simply, in the actual physics there is no such thing as a well-defined branch number. Similarly, in the actual physics there is no division of the dynamics into discrete branching events followed by evolution of individual branches: branching, rather, is continuous. But if branching is always going on, and cannot be quantified in a non-arbitrary manner, then no strategy can be formulated which is other than indifferent to the presence of branching.

A quick defence of state supervenience would be: the agent's preferences supervene on the actual state of the branch; transformations which differ only in how they would affect non-actual quantum states do not differ in any relevant respect. 

\begin{description}
\item[Skeptic:]Hang on. This brings out a tacit assumption in the formalism you've adopted:  the idea that acts can be represented by \emph{single} unitary transformations rather than by \emph{sequences} of unitary transformations. Why regard a sequence of measurements as decision-theoretically equivalent to a single measurement just because the same unitary transformation is enacted by both?
\item[Author:] Here's one possible defence. The agent is playing a sequence of games which result in rewards that he spends only after the sequence is done. In this case, what does he care about what happens during the brief period in which the games are being played (when having or not having rewards makes no difference to his status) --- should he not care only about the state of the universe after the payouts are all made?
\item[Skeptic:]Well, that sounds intuitive, but so what? We're discussing \emph{the Everett interpretation} --- appeals to intuition are going to ring a little hollow here.
\item[Author:]Fair enough. A far better defence is to observe that caring about the final state only is the diachronic equivalent of branch indifference, and can be defended in the same way. There is no ``real'' branching structure beyond a certain fineness of grain, so the details of that structure can only be included in terms of their coarse-grained consequences. 

Put another way: we could have defined our decision theory in terms of preferences, not over final states, but over consistent history spaces. But if we had done so, we would have needed both  synchronic and diachronic indifference assumptions: indifference both to the fineness of grain of the history projectors at each time, and to the size of the temporal gaps between history projectors. Translated back into our setting, where we consider sequences of decisions made only over very short periods of time, the former assumption entails branch indifference and the latter entails that acts can be represented by single unitary transformations.
\end{description}

\section{A quantum representation theorem}

We can now prove, in succession, three results, the first three of which are (trivially) entailed by the fourth.
\begin{description}
\item[Equivalence lemma:] If two acts assign the same weight to each reward, the agent must be indifferent between them.
\item[Nullity lemma:] An agent is indifferent to a possible outcome of an act iff that act has weight zero.
\item[Dominance lemma:] Suppose that two acts each only have two possible rewards $r_1$, $r_2$ as outcomes, with $r_1 \succ r_2$\footnote{That is: with an act which returns some microstate in $r_1$ with certainty preferred to one which returns some microstate in $r_2$ with certainty; that this determines a well-defined ordering over rewards follows from microstate indifference.} and that the first act assigns a higher weight to $r_1$ than the second act does. Then the first act must be preferred to the second.
\item[Born rule theorem:] There is a utility function on the set of rewards, unique up to affine transformations, such that one act is preferred to another iff its expected utility, calculated with respect to this utility function and to the quantum-mechanical weights of each reward, is higher.
\end{description}
Since all these results are proved \emph{formally} in section~\ref{wallace-formalproof}, my purpose in this section is explanation and not persuasion: I wish simply to show the general shape of the proof.

The equivalence lemma is best illustrated by examples (here I basically follow the argument of \citeN{wallaceprobdec}). For a simple case, suppose we have two acts (A and B, say): in each, a system is prepared in a linear superposition $\alpha \ket{+}+\beta \ket{-}$ and then measured in the $\{\ket{+},\ket{-}\}$. On act A, a reward is then given if the result is `+'; on B, the same reward is given on `-' instead. The resultant states are
\be
\mathrm{A:}\,\,\,\,\alpha\tpk{+}{\mbox{reward}}+\beta\tpk{-}{\mbox{no reward}};
\ee
\be
\mathrm{B:}\,\,\,\,\alpha\tpk{+}{\mbox{no reward}}+\beta\tpk{-}{\mbox{reward}}.
\ee
By erasure, there will exist acts available to the agent's future self in the reward branch (for both A and B) which erase the result of what was measured, leaving only the reward. Performing these transformations, and the equivalent erasures in the no-reward branch, leaves
\be
\mbox{A-plus-erasure:}\,\,\,\,\alpha\tpk{0}{\mbox{reward}}+\beta\tpk{0'}{\mbox{no reward}};
\ee
\be
\mbox{B-plus-erasure:}\,\,\,\,\beta\tpk{0}{\mbox{reward}}+\alpha\tpk{0'}{\mbox{no reward}}.
\ee
Now, by branch indifference, the agent's future selves are indifferent to whether this 
erasure is or is not performed. (Branch indifference is needed because we have no guarantee that erasures are non-branching; if we did, microstate indifference would suffice). So by diachronic consistency, the original agent is indifferent between A and A-plus-erasure, and between B and B-plus-erasure.

But now: if $\alpha=\beta$, then A-plus-erasure and B-plus-erasure leave the system in the same quantum state. So by state supervenience, the agent is indifferent between them. Since we know from ordering that preferences are transitive, the agent must also be indifferent between A and B. Indeed, we actually require only that  $|\alpha|=|\beta|$, for phase differences too can be erased.

For a slightly more complicated case, suppose game C involves a 2-state system being prepared in state \[\sqrt{2/3}\ket{+}+\sqrt{1/3}\ket{-}\] and a reward being given on `+', and game D involves a 3-state system being prepared in state \[\sqrt{1/3}(\ket{+}+\ket{0}+\ket{-})\] and a reward being given on `+' and on `0'. The resultant states are then
\be
C:\sqrt{2/3}\tpk{+}{\mbox{reward}}+\sqrt{1/3}\tpk{-}{\mbox{no reward}};
\ee
\be
D:\sqrt{1/3}\tpk{+}{\mbox{reward}}+\sqrt{1/3}\tpk{0}{\mbox{reward}}+\sqrt{1/3}\tpk{-}{\mbox{no reward}}.
\ee
But by erasure, there is an act available for the future self of the agent in the `reward' branch of game C which creates two equally-weighted branches:
\be
\tpk{+}{\mbox{reward}}\longrightarrow \sqrt{1/2}\tpk{X}{\mbox{reward}}+\sqrt{1/2}\tpk{Y}{\mbox{reward}}
\ee
Since by branch indifference the agent's future self is indifferent to performing this act or not, by diachronic consistency the original agent is indifferent between C and C-plus-branching. But the state produced by C-plus-branching is 
\be
\mbox{C-plus-branching}:\sqrt{1/3}\tpk{X}{\mbox{reward}}+\sqrt{1/3}\tpk{Y}{\mbox{reward}}+\sqrt{1/3}\tpk{-}{\mbox{no reward}}.
\ee
By a generalisation of our earlier argument, the agent is indifferent between C-plus-branching and D, and so between C and D.

By arguments of this kind, the equivalence lemma can be proved for any act with finitely many outcomes. The null and dominance lemmas are easy further steps, using the second clause of diachronic consistency.

We are now nearly done: the remainder of the proof is actually a standard decision-theoretic method for constructing utilities. Pick two rewards $R$ and $S$ with $R\succ S$, and assign $R$ utility 1 and $S$ utility 0. For any reward $T$ satisfying $R \succeq T \succeq S$, there is a unique number $U(T)$ such that the agent is indifferent between getting $T$ with certainty, and getting $R$ on a branch of weight $U(T)$ and $S$ otherwise. (We need continuity to establish this and rule out the possibility of rewards whose utilities differ only infinitesimally.)

Now consider an act which leads to rewards $R,S,T$ with weights w(R), w(S) and w(T) respectively. The agent's future selves in the $T$ branch are indifferent between doing nothing and performing an act that delivers $R$ with weight $U(T)$ and $S$ otherwise. Applying diachronic consistency once more, the original agent is indifferent between the original act and an act which delivers $R$ with weight $w(R)+w(T)U(T)$ and an act which delivers $S$ with weight $w(S)+(1-U(T))w(T)$. Note that the utilities of these acts are the same: in this particular case, the agent is indifferent between two acts iff they have the same utility. Generalising the argument, and applying the dominance lemma, tells us that one act is preferred to another iff its utility is higher.

The continuity axioms play only a limited role in these arguments. They serve to rule out situations where two rewards are infinitesimally, or infinitely, different in value; they are also required to handle the generalisation to acts which have infinitely many rewards as possible outcomes.

\section{Formal statement of the axioms}\label{wallace-formalstatement}

As promised, in this section and the next I lay out the formal version of my decision theory and its associated proofs. The reader who is happy to take on trust my mathematics --- and my reassurances that there has been no sleight of hand --- is welcome to skip to section~\ref{wallace-otherstrategies}.

A \emph{quantum decision problem} is specified by:
\begin{itemize}
\item A separable Hilbert space \mc{H}. Given a set $\mc{S}$ of subspaces of $\mc{H}$, I write $\vee \mc{S}$ (the \emph{disjunction} of $\mc{S})$ for the closure of the span of $\cup \mc{S}$, and $\wedge \mc{S}$ (the \emph{conjunction} of \mc{S}) for the closure of $\cap \mc{S}$; Given subspaces $E$ and $F$, I define $E \vee F=\vee\{E,F\}$ and likewise for $\wedge$, and I write $\Pi_E$ for the projector onto $E$.

\item A complete Boolean algebra $\mc{E}$ of subspaces of $\mc{H}$, the \emph{event space}. (So $\mc{E}$ contains $\mc{H}$ and is closed under $\vee$, $\wedge$, and taking the complement.) I define a \emph{partition} of an event $E$ to be a set of mutually orthogonal events whose conjunction is $E$.

\item A subset $\mc{M}$ of $\mc{E}$, the \emph{macrostates}, such that for any event $E$, there is a partition of $E$ by macrostates. 

\item For each $E \in \mc{E}$, a set $\mc{U}_E$ of unitary operators from $E$ into \mc{H}, which we call the  set of \emph{acts available at }$E$. We write $\mc{O}_U$ for the smallest event containing the range of the act $\opm{U}$\footnote{We can define $\mc{O}_U$ explicitly as the conjunction of all events containing the range of $\opm{U}$; this suffices to show that $\mc{O}_U$ is well-defined.} and require that the choice of available acts satisfies:
\begin{enumerate}
\item \emph{Restriction}: If $E,F\in \mc{E}$ and $F\subset E$, then if $\opm{U}$ is available at $E$ then the unitary map $\opm{U}|_F$, defined by  $\opm{U}\psi=\opm{U}|_F\psi$ whenever $\psi \in F$, is available at $F$.
\item \emph{Composition}: If $\opm{U}$ is available at $E$, and $\opm{V}$ is available at  $\mc{O}_U$, then $\opm{V}\opm{U}$ is available at $E$.
\item \emph{Indolence}: For any event $E$, if there are any acts available at $E$ then the identity $\id_E$ is available at $E$. (More precisely, the embedding map of $E$ into \mc{H} is available at $E$.)
\item \emph{Continuation}: If $\opm{U}$ is available at some $E$, then there is some act available at $\mc{O}_U$.
\item \emph{Irreversibility}: If $\opm{U}$ is available at $E\vee F$, $\mc{O}_{U|_E}\wedge \mc{O}_{U|_F}=\emptyset$.
\end{enumerate}
\item A partition \mc{R} of \mc{E} (that is, a set of mutually orthogonal elements of \mc{E} whose disjunction is \mc{H}), the set of \emph{rewards}. These represent payoffs an agent could get.
\end{itemize}
The simplest choice of macrostates and event space is to pick some particular set of orthogonal subspaces of \mc{H} whose disjunction is \mc{H}, take this as \mc{M}, and take \mc{E} to be the set of all disjunctions of subsets of \mc{M}; this is the sense of ``macrostate'' and ``event'' used in the informal version of the proof. However, we could equally well take \mc{E} to be an arbitrary Boolean algebra of subspaces and define $\mc{E}=\mc{M}.$ (As was noted previously, this sort of formalisation might be more appropriate for decision problems with a less natural discrete structure.)

Rays within $\mc{H}$, as usual, are called states. I adopt the usual convention of representing a ray by any vector within it and of blurring the distinction between the two; I do not require that vectors representing states be normalised. (This is just for notational convenience.) If $\mc{B}(E,\mc{H})$ is the set of unitary maps from $E$ into \mc{H}, it can naturally be regarded as a subset of $\mc{B}(\mc{H},\mc{H})$ by identifying $\opm{U}$ with $\opm{U}\Pi_E$; as such, $\mc{B}(E,\mc{H})$ inherits the norm topology.

I introduce a few derived concepts. The \emph{weight} $\mc{W}_\psi(E|\opm{U})$ of an event $E$ with respect to a state $\psi$ and an act $\opm{U}$  is defined by 
\be
 \mc{W}_\psi(E|\opm{U})=\| \Pi_{E}\opm{U}\ket{\psi}\|^2 =\matel{\psi}{\opmad{U}\Pi_{E}\opm{U}}{\psi}.
\ee
A \emph{reward function} is any function from \mc{R} to $[0,1]$ such that $\sum_{r\in \mc{R}} w(r)=1$. Any pair of a state $\psi\in E$ and an act $\opm{U}$ available at $E$ determines a reward function 
\be 
R_{\psi,U}(r)=\mc{W}_\psi(r|\opm{U})
\ee 
which I call the \emph{characteristic reward function} of $\opm{U}$ and $\psi$.

A set $\mc{F}$ of events is \emph{available} if they are mutually orthogonal and there is at least one act available at $\vee \mc{F}$. (An event is available iff its singleton set is available).

Finally, if \mc{S} is any set of rewards, I say that an act $A$ \emph{has rewards in} \mc{S} iff its range is a subset of $\vee \mc{S}$. If  $u$ is a real function of $\mc{S}$ and $\opm{U}$ is an act whose rewards are in \mc{S}, the \emph{expected utility} of \opm{U} with respect to a state $\psi$ (and, tacitly, with respect to $u$) is
\be
\EU_\psi(\opm{U})=\sum_{r\in\mc{S}}\mc{W}_\psi(r|\opm{U})u(r)\equiv \sum_{r\in\mc{S}}R_{\psi,U}(r)u(r).
\ee

Stating the richness axioms is a little fiddly, because of the need to make sure not only that certain acts (erasures, branchings etc) are available everywhere, but to make sure that they are available on multiple branches concurrently. To state them in a concise way, I make the following definitions. Firstly, if $\mc{P}=\{p_1,p_2, \ldots \}$ is a (countable or finite) set of positive real numbers whose sum is unity, and $\psi\in M\subset r$ for some state $\psi$, macrostate $M$, and reward $r$, then a $\mc{P}$-branching of $\psi$ is some act $\opm{U}$ available at $M$ such that $\mc{O}_U\subset r$ and such that there is a partition $\mc{M}=\{M_1,M_2,\ldots\}$ of $\mc{O}_U$ by macrostates with $\mc{W}_\psi(M_i|\opm{U})=p_i$. (Informally, a $\mc{P}$-branching is an act which splits the agents branch into many branches, each having the same weight as an element of $\mc{P}$, but without changing the rewards that the agent gets.)

Secondly, if $M$ and $M'$ are macrostates with $M\subset r$ and $M' \subset r$ for some reward $r$, and $\psi,\psi'$ are states in $M$, $M'$ respectively, then an \emph{erasure} of $\psi$ and $\psi'$ is a pair of acts $\opm{U}$, $\opm{U}'$ available at $M$ and $M'$ respectively, such that $\mc{O}_U$ and $\mc{O}_{U'}$ are both subsets of $r$ and $\opm{U}\psi=\opm{U'}\psi'$.

And thirdly, if $\mc{F}$ is an available set of events, an \emph{act function} $\mc{U}$ for that set is a function which assigns to each $F\in\mc{F}$ an act $\mc{U}(F)$ available at $F$. An act function is \emph{compatible} if 
\be 
\sum_{F\in\mc{F}}\mc{U}(F)\Pi_F
\ee
is available at $\vee \mc{F}$.

The richness axioms are now stateable:
\begin{description}
\item[Reward availability:] 
Suppose that $\mc{F}$ is an available set of macrostates and $f$ is a function from $\mc{F}$ into rewards. 

Then there is a compatible act function $\mc{U}$ for $\mc{F}$ with $\mc{U}(F)\subset f(F)$ for all $F\in \mc{F}$.
\item[Branching availability:] 
Suppose that $\mc{F}$ is an available set of macrostates and for each $F\in \mc{F}$, $\psi_F$ is a nonzero state in $F$ and $\mc{P}_F$ is a (finite or countable) set of positive real numbers summing to unity.

Then there is a compatible act function $\mc{U}$ for $\mc{F}$ such that for each $F\in\mc{F}$, $\mc{U}(F)$ is a $\mc{P}_F$-branching of $\psi_F$.

\item[Erasure:] Suppose that $\{r_1,r_2,\ldots\}$ is a (finite or countable) set of rewards, that $\mc{M}=\{M_1,M_2,\ldots\}$ and $\mc{N}=\{N_1,N_2,\ldots\}$ are two available sets of macrostates with $M_i\subset r_i$ and $N_i\subset r_i$, and that for each $i$, $\psi_i\in M_i$ and $\varphi_i\in N_i$ are nonzero states.

Then there are compatible act functions $\mc{U}$ for $\mc{M}$ and $\mc{V}$ for $\mc{N}$ such that, for each $i$, $(\mc{U}(M_i),\mc{V}(N_i))$ is an erasure of $\psi_i$ and $\varphi_i$.
\item[Problem Continuity:] For every available $E$, the set of acts available at $E$ is an open subset (in operator norm topology) of the set of unitary maps from $E$ to \mc{H}.\footnote{The operator norm topology on the set of linear maps between normed spaces $V$ and $W$ is defined by the norm $\|U\|=\sup \{\|Ux\|:\|x\|=1\}$. The set of unitary maps from $E$ to \mc{H} is a subset of the set of all maps between those two spaces, and inherits the latter's topology.}
\end{description}
Notice that reward availability and preparation together entail that for any reward function and any $\psi\in E$, there is an act \opm{U} available at $E$ such that $\psi$ and $\opm{U}$ have that reward function as their characteristic reward function. 

We now define a \emph{state-dependent solution} to a decision problem as specified by  an assigment to every available macrostate $E$, and every state $\psi\in E$ , of a two-place relation $\succ^\psi$ on the acts available at $E$. (Strictly our notation should include $E$ but for simplicity, its value will always be tacit.) 

We call an event $N$ \emph{null} for a given state $\psi$ and act \opm{U} iff, whenever acts $V_1$ and $V_2$ are identical on the complement of $N$, $\opm{V}_1\opm{U}\sim^\psi \opm{V}_2\opm{U}$. (So an event is null if the agent doesn't care what happens to his future selves, if any, in the branch defined by that event. We will shortly see that, as expected, an event is null iff there are in fact no such future selves.) It is easy to see that any finite union of null sets is null, as is any subset of a null set.

We can now state the rationality axioms:
\begin{description}
\item[Ordering:] For every $\psi$ for which it is defined, $\succ^\psi$ is a total ordering. That is: it is transitive, asymmetric, and the relation $\sim^\psi$, defined by $E\sim^\psi F$ iff neither $E\succ^\psi F$ nor $F \succ^\psi E$, is an equivalence relation. (As usual, we write `$E \succeq^\psi F$' as an abbreviation for `either $E\succ^\psi F$ or $E\sim^\psi F$'.)
\item[Diachronic Consistency:] Suppose $\opm{U}$ is available at $E$, and 
$\opm{V}_1$ and $\opm{V}_2$ are available at $\mc{O}_U$. Then:
\begin{enumerate}
\item[(i)] If there is some partition \mc{P} of $\mc{O}_U$ into macrostates such that  $\opm{V}_1|_E\succeq^{\Pi_E\opm{U}\psi}\opm{V}_2|_E$ for every element $E$ of the partition  not null with respect to $\psi$ and $\opm{U}$, then $\opm{V}_1\opm{U}\succeq^\psi\opm{V}_2\opm{U}$.
\item[(ii)] If in addition,  $\opm{V}_1|_E\succ^{\Pi_E\opm{U}\psi}\opm{V}_2|_E$ for at least one such $E$, then $\opm{V}_1\opm{U}\succ^\psi\opm{V}_2\opm{U}$.
\end{enumerate}
\item[Macrostate indifference:] If: 
\begin{itemize}
\item $\opm{U},\opm{V}$ are acts available at $M$;
\item $\opm{U}',\opm{V}'$ are acts available at $M'$;
\item $\mc{O}_U\subset M_1\wedge r_1$ and $\mc{O}_{U'}\subset M_1\wedge r_2$  for some macrostate $M_1$ and reward $r_1$;
\item $\mc{O}_V\subset M_2\wedge r_2$ and $\mc{O}_{V'}\subset M_2\wedge r_2$ for some macrostate $M_2$ and reward $r_2$
\end{itemize}
then for any $\psi,\psi '$ with $\psi\in M$ and $\psi'\in M'$, $\opm{U}\succeq^\psi\opm{V}$ iff $\opm{U'}\succeq^{\psi'}\opm{V'}$.
\item[Branching indifference:] If:
\begin{itemize}
\item $r$ is a reward;
\item $M$ is a macrostate with $M\subset r$;
\item $\opm{U}$ is available at $M$;
\item $\psi\in M$ and $\opm{U}\psi\in r$
\end{itemize}
then $\opm{U}\sim^\psi \id_M$.
\item[State supervenience:]
If:
\begin{itemize}
\item $\psi\in E$ and $\psi'\in E'$ for macrostates $E,E'$;
\item $\opm{U}$ and $\opm{V}$ are available at $E$, and $\opm{U'}$ and $\opm{V'}$ are available at $E'$;
\item $\opm{U}\psi=\opm{U}'\psi'$ and $\opm{V}\psi=\opm{V}'\psi'$
\end{itemize}
then $\opm{U}\succ^\psi\opm{V}$ iff $\opm{U'}\succ^{\psi'}\opm{V'}$.
\item[Solution continuity:]
If $E$ is a macrostate and $\psi\in E$, if $\opm{U},\opm{U}'$ are available at $E$, and if $\opm{U}\succ^\psi \opm{U}'$, then in the space of unitary maps from $E$ into \mc{H} there are  neighbourhoods (in norm topology) $\mc{N},\mc{N}'$ of $\opm{U}$, $\opm{U}'$ respectively such that any act in $\mc{N}$ available at $E$ is preferred (at $\psi$) to any act in $\mc{N}'$ available at $E$.
\end{description}

Given a solution to a quantum decision problem, we can use it to define a preference ordering on rewards: for any two rewards, $r_1\succ r_2$ iff there is some macrostate $E$, some state $\psi\in E$,  and acts $\opm{U}_1,\opm{U}_2$ available at $E$ such that $\mc{O}_{U_i}\subset r_i$ and $\opm{U}_1 \succ^\psi \opm{U}_2$. Provided that the problem is reward-available and the solution is macrostate-indifferent and branching-indifferent, this preference order is a total ordering on \mc{R}. If $r$ and $s$ are rewards with $r\preceq s$, I will say that a reward $t$ is between $r$ and $s$ iff $s \succeq t \succeq t$; I write $[r,s]$ for the set of rewards between $r$ and $s$.

If \mc{M} consists of some set of orthonormal subspaces (as in the informal proof), then this observation more or less exhausts the usefulness of macrostate indifference. At the other extreme, if \mc{M}=\mc{E} then macrostate indifference actually entails branch indifference. The distinction between the axioms, then, is a matter of how we mathematically represent the branching structure --- which is appropriate, since the motivation for branching indifference itself is that the details of that structure are an unphysical artefact of the mathematics.

(The mathematically inclined reader may be wondering at this point if the axioms are consistent. To show that they are, consider the following model. Let $\mc{H}_R$ be a two-dimensional Hilbert space with an orthogonal basis $\{\ket{+},\ket{-}\}$; for each $N>0$ let $\{\mc{H}_N\}$ be an $N$-dimensional Hilbert space with an orthonormal basis $\{\ket{N,1},\ket{N,2},\ldots \ket{N,N}\}$. 

Now: take the Hilbert space of our decision problem to be
\begin{equation}
\mc{H}=\mc{H}_R \otimes \left( \oplus_{I=1}^\infty \mc{H}_I   \right),
\end{equation}
so that a complete basis of states is 
\be 
\tpk{\pm}{N,M} \,\,\,(M\leq N),
\ee 
and take the macrostates to consist of all the one-dimensional subspaces spanned by each of these states, and the events to be all disjunctions of macrostates.
 The available events are all those which are contained in some fixed $\mc{H}_R\otimes \mc{H}_N$, and the acts available at an available event contained in $\mc{H}_R\otimes \mc{H}_N$ are all unitary maps from $\mc{H}_R\otimes \mc{H}_N$ to $\mc{H}_R\otimes \mc{H}_{N'}$, with $N'>N$. The reward subspaces are $\mc{H}^{\pm}=\{\mathrm{Span}\,\ket{\pm} \}\otimes \mc{H}$. Finally, 
 an act $\opm{U}$ is preferred to an act $\opm{U}'$ at \ket{\psi} iff
\be
\left\|(\proj{+}{+}\otimes \id)\opm{U}\ket{\psi}\right\|>\left\|(\proj{+}{+}\otimes \id)\opm{U}'\ket{\psi}\right\|.
\ee
I leave readers to satisfy themselves that this system does indeed obey the axioms; the preference order is, of course, the Born rule.)

\section{Formal statement and proof of the representation theorem}\label{wallace-formalproof}
\begin{description}
\item[Equivalence Lemma:] Suppose that:
\begin{enumerate}
\item[(i)] \mc{P} is a quantum decision problem satisfying erasure, branch availability and reward availability;
\item[(ii)] $\succ^\psi$ is a state-dependent solution to $\mc{P}$ satisfying ordering,  diachronic consistency, macrostate indifference, branching indifference, and state supervenience;
\item[(iii)]$\opm{U}$ and $\opm{V}$ are available at $E$, and $\opm{U'}$ and $\opm{V'}$ are available at $E$';
\item[(iv)] $\psi\in E$ and $\psi'\in E'$;
\item[(v)] $R_{\psi,U}=R_{\psi',U'}$ and  $R_{\psi,V}=R_{\psi',V'}$.
\item[(vi)] The reward functions of the acts are each non-zero for only finitely many rewards.
\end{enumerate}
Then $\opm{U}\succ^\psi\opm{V}$ iff $\opm{U}'\succ^{\psi'}\opm{V}'$.

\end{description}
\Proof For each reward $r$ for which $R_{\psi,U}(r)\neq 0$, let $\mc{M}_r$ and $\mc{N}_r$ be partitions of $\mc{O}_U\wedge r$ and $\mc{O}_{U'}\wedge r$ respectively, and let $\#M_r$ and $\#N_r$ be the number of elements (finite or infinite) in $\mc{M}_r$ and $\mc{N}_r$ respectively.

Define the sets $\mc{P}_r$ (for each $r$)
\begin{equation}
\mc{P}_r=\{\mc{W}_{\psi'}(N|\opm{U}')/\mc{W}_{\psi'}(r|\opm{U}'):N\in\mc{N}_r\}
\end{equation}
These  are sets of positive real numbers summing to unity, so by branching availability there is an act $\opm{W}$ available at $\mc{O}_U$ such that, for each $r$ and each $M\in \mc{M}_r$, $\opm{W}|_M$ is a $\mc{P}_r$-branching of $\Pi_M\opm{U}\psi$: it splits $\Pi_M\opm{U}\psi$, which has weight $\mc{W}_\psi(M|\opm{U})$, into $\#\mc{N}_r$ states, one for each $N\in\mc{N}_r$, with weights $\mc{W}_\psi(M|\opm{U})\times \mc{W}_{\psi'}(N|\opm{U'})/\mc{W}_{\psi'}(r|\opm{U}')$. There is therefore\footnote{We appeal here to the irreversibility requirement on decision problems.} a partition $\mc{W}$ of $\mc{O}_{W}$ into macrostates, such that:
\begin{itemize}
\item For each reward $r$ there are $\#\mc{M}_r\times\#\mc{N}_r$ elements of $\mc{W}$ in $r$.
\item Each such element can be labelled by pairs of elements from $\mc{M}_r$ and $\mc{N}_r$: let us write it as $K^r_{M,N}$.
\item $\mc{W}_\psi(K^r_{M,N}|\opm{W}\opm{U})=
\mc{W}_\psi(M|\opm{U})\times \mc{W}_{\psi'}(N|\opm{U'})/\mc{W}_{\psi'}(r|\opm{U}').$
\end{itemize}
Furthermore, by branching indifference, $\opm{W}|_M\sim^{\Pi_M\opm{U}\psi}\id_M$ for any macrostate $M$, and hence by diachronic consistency, $\opm{W}\opm{U}\sim^\psi\opm{U}$.

Applying the same procedure with $\opm{U}$ and $\opm{U}'$ reversed yields an act $\opm{W}'$ such that $\opm{W}'\opm{U}\sim^\psi \opm{U}$, and a partition $\mc{W}'$ of $\mc{O}_{W'}$ by macrostates, such that 
\begin{itemize}
\item For each reward $r$ there are $\#\mc{M}_r\times\#\mc{N}_r$ elements of $\mc{W}'$ in $r$.
\item Each such element can be labelled by pairs of elements from $\mc{M}_r$ and $\mc{N}_r$: we write it as $K'^r_{M,N}$.
\item $\mc{W}_{\psi'}(K'^r_{M,N}|\opm{W}\opm{U})=
\mc{W}_\psi(M|\opm{U})\times \mc{W}_{\psi'}(N|\opm{U'})/\mc{W}_{\psi}(r|\opm{U}).$
\end{itemize}
But since
 \be\mc{W}_{\psi}(r|\opm{U})\equiv \mc{R}_{\psi,U}(r)=\mc{R}_{\psi',U'}(r)\equiv\mc{W}_{\psi'}(r|\opm{U}'),\ee
it follows that $\mc{W}_\psi(K^r_{M,N}|\opm{W}\opm{U})=\mc{W}_{\psi'}(K'^r_{M,N}|\opm{W}'\opm{U}').$

So we have constructed acts $\opm{W}$, $\opm{W}'$ and partitions $\mc{W}=\{W_1, \ldots \}$, $\mc{W}'=\{W'_1, \ldots\}$ of $\mc{O}_{W}$, $\mc{O}_{W'}$ by macrostates such that:
\begin{enumerate}
\item For any $i$, $W_i$ exists iff $W'_i$ does (\iec, the two partitions have the same number of elements) and there is some reward $r$ such that $W_i$ and $W'_i$ are elements of $r$. 
\item $\mc{W}_\psi(W_i|\opm{W} \opm{U})=\mc{W}_\psi(W'_i|\opm{W}'\opm{U})$ for all $W_i$.
\end{enumerate}
Now define 
\be\chi_i=\Pi_{W_i}\opm{W}\opm{U}\psi/\|\Pi_{W_i}\opm{W}\opm{U}\psi\|\ee
and
\be\chi'_i=\Pi_{W'_i}\opm{W'}\opm{U'}\psi'/\|\Pi_{W'_i}\opm{W'}\opm{U'}\psi'\|.\ee
By erasure, there exist acts $\opm{X}$, $\opm{X}'$ available at $\mc{O}_W$, $\mc{O}_{W'}$ such that $(\opm{X}|_{W_i})\chi_i=(\opm{X}'|_{W'_i})\chi'_i$. By branching indifference, $\opm{X}|_{W_i}\sim^{\chi_i}\id_{W_i}$, so by diachronic consistency, $\opm{X}\opm{W}\opm{U}\sim^\psi\opm{W}\opm{U}\sim^\psi \opm{U}$; similarly, $\opm{X'}\opm{W'}\opm{U'}\sim^{\psi'}\opm{U}'$.

Since 
\be
\opm{X}\opm{W}\opm{U}\psi=\sum_i \mc{W}_\psi(W_i|\opm{W} \opm{U}) (\opm{X}|_{W_i})\chi_i,
\ee
it follows that $\opm{X}\opm{W}\opm{U}\psi=\opm{X'}\opm{W'}\opm{U'}\psi'$.

So: for \opm{U} and \opm{U'}, we have found acts $\opm{Y}=\opm{X}\opm{W}\opm{U}$ and $\opm{Y'}=\opm{X'}\opm{W'}\opm{U}'$ such that $\opm{U}\sim^\psi\opm{Y}$, $\opm{U'}\sim^{\psi'}\opm{Y'}$, and $\opm{Y}\psi=\opm{Y'}\psi'$. Repeating this process for \opm{V} and \opm{V'}, we can find acts $\opm{Z},\opm{Z'}$ such that $\opm{Z}\sim^\psi\opm{V}$, $\opm{Z'}\sim^{\psi'}\opm{V'}$, and $\opm{Z}\psi=\opm{Z'}\psi'$. The conclusion now follows immediately from state supervenience.
$\Box$

Because of the equivalence lemma, there is a unique total ordering defined on the set of all reward functions, which we once again write as $\succ$ (note that it is state-independent).

\begin{description}
\item[Nullity Lemma:] Suppose that:
\begin{enumerate}
\item[(i)] \mc{P} is a quantum decision problem satisfying erasure, branch availability and reward availability;
\item[(ii)] $\succ^\psi$ is a state-dependent solution to $\mc{P}$ satisfying ordering,  diachronic consistency, macrostate indifference, branching indifference, and state supervenience;
\item[(iii)] There exist rewards $r,s$ with $r\succ s$.
\end{enumerate}
Then an event $E$ is null with respect to a state $\psi$ and an act $\opm{U}$ iff $\matel{\psi}{\opmad{U}\Pi_E\opm{U}}{\psi}=0$.
\end{description}
\Proof Let $\matel{\psi}{\opmad{U}\Pi_E\opm{U}}{\psi}=\alpha.$ An event is null if and only if, given acts $\opm{V}$ and $\opm{W}$ available at $\mc{O}_U$ which are identical except on $E$, $\opm{V}\opm{U}\sim^\psi\opm{W}\opm{U}$. Given the Equivalence Lemma, any two such acts are equivalent whenever they have the same weight function, so if $E$ is null for $\psi$ and $\opm{U}$, any event $E'$ is null with respect to some $\opm{U}'$ and $\psi'$ whenever 
$
\matel{\psi'}{\opmad{U'}\Pi_E'\opm{U'}}{\psi'}=\alpha.
$
If $\alpha>0$, then $\alpha>1/N$ for some $N$. By combining branch availability with reward availability, we can construct some act $\opm{V}$ and state $\varphi$  with weight function 
\begin{eqnarray*}
\mc{W}_\varphi(E_1|\opm{V})&=&1/N\\
\mc{W}_\varphi(E_2|\opm{V})&=&\alpha-1/N\\
\mc{W}_\varphi(E_3|\opm{V})&=&1-\alpha\\
\end{eqnarray*}
$E_1\vee E_2$ is null (wrt $\varphi$ and $\opm{V}$), hence $E_1$ is, hence any event with weight $1/N$ is. Applying branch availability and reward availability  again, we can find $\varphi'$, $\opm{W}$ and $F_1,\ldots F_N$ such that $\mc{W}_{\varphi'}(F_i|\opm{W})=1/N$. Each $F_i$ is null wrt $\varphi'$ and $\opm{W}$, hence so is $\mc{E}$. This contradicts premise (iii), since if all events are null then all rewards are equivalent. 

Conversely, suppose that some event has weight zero. Its nullity now follows from state supervenience, since no change to the physical state is enacted by any transformation restricted to that event.
$\Box$

\begin{description}
\item[Dominance Lemma:] Suppose that 
\begin{itemize}
\item[(i)] \mc{P} is a quantum decision problem satisfying erasure, branch availability and reward availability;
\item[(ii)] $\succ^\psi$ is a state-dependent solution to $\mc{P}$ satisfying ordering,  diachronic consistency, macrostate indifference, branching indifference, and state supervenience;
\item[(iii)] $s,t$ are rewards with $s\succ t$.
\item[(iv)] $f[\alpha]$ is the reward function defined by $f[\alpha](s)=\alpha$, $f[\alpha](t)=1-\alpha$, $f[\alpha](r)=0$ for all other $r$.
\end{itemize}
Then $f[\alpha]\succ f[\beta]$ iff $\alpha>\beta$.
\end{description}
\Proof This is an easy corollary of the Nullity Lemma. Suppose $\alpha>\beta$, then by branch availability and reward availability, there will be some act $\opm{A}$ and state $\varphi$  with weight function 
\begin{eqnarray*}
\mc{W}_\varphi(E_1|\opm{A})&=&\beta\\
\mc{W}_\varphi(E_2|\opm{A})&=&\alpha-\beta\\
\mc{W}_\varphi(E_3|\opm{A})&=&1-\alpha\\
\end{eqnarray*}
By reward availability there exist sets of compatible acts $\{\opm{U}_1,\opm{U}_2,\opm{U}_3\}$ and $\{\opm{V}_1,\opm{V}_2,\opm{V}_3\}$ such that $\opm{U}_i$ and $\opm{V}_i$ are available at $E_i$, and 
such that $\opm{U}_1,\opm{V}_1$ and $\opm{U}_2$ have outcomes all lying in $s$ and $\opm{V}_2,\opm{U}_3$ and $\opm{V}_3$ have outcomes all lying in $t$. By macrostate indifference and branching indifference 
$\opm{U}_{i}\simeq^\chi\opm{V}_{i}$ for any $\chi \in E_i$
and in particular 
$\opm{U}_{2}\succ^\chi\opm{V}_{2}$ for any $\chi \in E_2$.

If we define 
\be
\opm{W_\alpha}=\opm{U}_1\Pi_{E_1}+\opm{U}_{2}\Pi_{E_2}+\opm{U}_3\Pi_{E_3}
\ee
and
\be
\opm{W_\beta}=\opm{V}_1\Pi_{E_1}+\opm{V}_{2}\Pi_{E_2}+\opm{V}_3\Pi_{E_3}
\ee
then by diachronic consistency, since $E_2$ is not null then $\opm{W_\alpha}\cdot \opm{A}\succ^\psi\opm{W_\beta}\cdot \opm{A}$. But the reward functions of $\opm{W_\alpha}\cdot \opm{A}$ and $\opm{W_\beta}\cdot \opm{A}$ are $f[\alpha]$ and $f[\beta]$ respectively, and the conclusion follows.
\begin{description}
\item[Utility Lemma:] Suppose that:
\begin{enumerate}
\item[(i)] \mc{P} is a quantum decision problem satisfying erasure, branch availability, and reward availability;
\item[(ii)] $\succ^\psi$ is a state-dependent solution to $\mc{P}$ satisfying ordering,  diachronic consistency, macrostate indifference, branching indifference, and state supervenience;
\item[(iii)] $s,t$ are rewards with $s\succ t$.
\item[(iv)] $u_s$, $u_t$ are real numbers with $u_s>u_t$.
\end{enumerate}
Then  there is a unique real function $u$ on  the set $[t,s]$ of rewards between $t$ and $s$  such that for any macrostate $E$, any state $\psi\in E$, and any two acts $\opm{U},\opm{V}$ available at $E$ whose rewards lie a finite subset of \mc{S},
\be
\opm{U}\succ_\psi\opm{V} \mbox{ whenever } \EU_\psi(\opm{U})>\EU_\psi(\opm{V})
\ee
(where the expected utilities are defined with respect to $u$, of course) and such that $u(s)=u_s$ and $u(t)=u_t$.
\end{description}
\Proof For simplicity we assume $u_s=1$ and $u_t=0$ (other values lead to a simple affine transformation of the utility function). We define the following reward functions: $f[\alpha]$ is defined as in the Dominance Lemma, and $g[r]$ is defined by $g[r](r')=\delta_{r,r'}.$

We now define $u(r)$ by
\be
u(r)=\mathrm{lub} \{\alpha: g[r]\succ f[\alpha ]\}.
\ee
Let $\{u_n(r)\}$ be a sequence of functions such that $u_m(r)\leq u(r)$ and $\lim_{n\rightarrow \infty} u_n(r)=u(r)$, and let $\opm{U}$ be any act available at $E$ whose rewards lie in \mc{S}. We write $E_r$ for $\mc{O}_U\wedge r$ and $\chi_r$ for the normalised projection of $\psi$ onto $E_r$.

From branching availability and reward availability, for each $n$ we can find a compatible set of states $\{A_n(r):R_{\psi,U}(r)\neq 0\}$ such that $\opm{A}_n(r)$ is available at $E_r$ and $\opm{A}_n$ has reward function $f[u_n(r)]$; we define $\mc{A}_n=\sum_{r\in \mc{S}}A_n(r)\Pi_{E_r}$. By construction, $\id_{E_r}\succeq^{\chi_r}A_n(r)$ for all $r$ and $n$, so by diachronic consistency $\opm{U}\succeq^\psi\mc{A}_n\cdot\opm{U}$. 

By definition, the reward function of $\mc{A}_n\cdot\opm{U}$ (with respect to $\psi$) is $f[\lambda_n]$, where
\be
\lambda_n=\sum_{r\in \mc{S}}\mc{W}_\psi(r|\opm{U})u_n(r). 
\ee

So if $f[\opm{U}]$ is the reward function of $\opm{U}$ (with respect to $\psi$), we have established that $f[\opm{U}]\succeq f[\lambda_n]$, and hence by the Dominance lemma, $f[\opm{U}]\succeq f[\lambda]$ whenever $\lambda <\lambda_n$ for some $n$. Since  $u_n(r)\rightarrow u(r)$ for each $n$ and $r$, 
$\lambda_n\rightarrow \EU_\psi(\opm{U})$,and hence $f[\opm{U}]\succ f[\lambda]$ whenever $\lambda<\EU_\psi(\opm{U})$. Applying the same argument with a decreasing sequence, $f[\opm{U}]\prec f[\lambda]$ whenever $\lambda>\EU_\psi(\opm{U})$.

Now suppose that $\opm{U}$ and $\opm{V}$ are two such acts with $\EU_\psi(\opm{U})>\EU_\psi(\opm{V})$. Then for any $\alpha$ lying between the two expected utilities, there will exist an act $\opm{W}$ with reward function (wrt $\psi$) $f[\alpha]$. We have proved that $\opm{U}\succ^\psi\opm{W}$, and $\opm{W}\succ^\psi \opm{V}$, so it follows that $\opm{U}\succ^\psi\opm{V}$.

To see that this utility function is unique, note that if there were another utility function $u'$ we could construct acts whose utilities were the same as calculated by this second utility, but not as calculated by the first; this contradicts the requirements on $u'$.
$\Box$

\begin{description}
\item[Born Rule Theorem:] Suppose that:
\begin{enumerate}
\item[(i)] \mc{P} is a quantum decision problem satisfying erasure, branch availability, reward availability and problem continuity;
\item[(ii)] $\succ^\psi$ is a state-dependent solution to $\mc{P}$ satisfying ordering,  diachronic consistency, macrostate indifference, branching indifference, state supervenience, and solution continuity.
\end{enumerate}
Then there is a function $u$ on the rewards of $\mc{P}$, unique up to positive affine transformations, such that if $\EU$ denotes the expected utility with respect to this function,
\be
\opm{U}\succ^\psi\opm{V} \mbox{ iff } \EU_\psi(\opm{U})>\EU_\psi(\opm{V}).
\ee
\end{description}
\Proof Note that problem continuity and solution continuity jointly entail that if $\opm{U}\succ^\psi\opm{U}'$, there are neighborhoods $\mc{N}, \mc{N}'$ of $\opm{U}$ and $\opm{U}'$ respectively such that all acts in $\mc{N}$ and $\mc{N}'$ are available and all acts in $\mc{N}$ are preferred (given $\psi$) to all acts in $\mc{N}'$. For simplicity I shall refer to this simply as continuity.

We begin by proving that if $s\succ r_1\succeq r_2\succ t$, then if the utilities determined by the Utility Lemma (via this choice of $s$ and $t$) for $r_1$ and $r_2$  coincide, then $r_1 \sim r_2$. Let this utility function be $u$ and again, for convenience take $u(s)=1$ and $u(t)$=0. Fix $E$ and $\psi\in E$, and let $\opm{U}_1$ and $\opm{U}_2$ be acts available at $E$ whose ranges lie in $r_1$ and $r_2$ respectively (by reward availability, some such acts exist). If $r_1\succ r_2$, then $\opm{U}_1\succ^\psi \opm{U}_2$. By continuity, there must exist neighborhoods $\mc{N}_1$, $\mc{N}_2$ of $\opm{U}_1$ and $\opm{U}_2$ such that any available act in $\mc{N}_1$ is preferred (given $\psi$) to any available act in $\mc{N}_2$. 

Now let $f_1[\alpha]$ and $f_2[\alpha]$ be reward functions with $f_1[\alpha](r_1)=1-\alpha$, $f_1[\alpha](t)=\alpha$ and
$f_2[\alpha](r_2)=1-\alpha$, $f_2[\alpha](s)=\alpha$. By branch availability and reward availability, there must exist some $\alpha$, and some acts $\opm{U}_{i,\alpha}$, such that $\opm{U}_{i,\alpha}\in \mc{N}_i$ and the reward function of $\opm{U}_{i,\alpha}$ (with respect to $\psi$) is $f_i[\alpha]$. 

So we have that $\opm{U}_{1,\alpha}\succ \opm{U}_{2,\alpha}$. But $\EU_\psi(\opm{U}_{1,\alpha})<\EU(\opm{U}_1)\equiv u(r_1)$, and $\EU_\psi(\opm{U}_{2,\alpha})>\EU(\opm{U}_2)\equiv u(r_2)$. So by the Utility lemma we must have that $u(r_1)>u(r_2)$.

We can now define a utility function for the whole of \mc{R}. For any rewards $r_1,r_2$ with $r_1\succ r_2$, and any real numbers $x_1$, $x_2$ with $x_1>x_2$, I will write $u[r_1,r_2,x_1,x_2]$ for the unique utility function determined on $[r_2,r_1]$ by setting the utility of $r_i$ to $x_i$.

Now, let $s,t$ be any two rewards with $s \succ t$ (if there are no such rewards, the theorem is true trivially). I define the utility of any reward $r$ by:
\begin{itemize}
\item If $s\succeq r \succeq t$, $u(r)=u[s,t,1,0](r)$.
\item If $r \succ s$, $u(r)$ is the unique value fixed by requiring that $u[r,t,u(r),0](s)=1.$
\item If $t \succ r$, $u(r)$ is the unique value fixed by requiring that $u[s,r,1,u(r)](s)=0.$
\end{itemize}
(Notice that this definition relies on the assumption that the utilities of $s$ and $t$ are guaranteed to be distinct.)

I now prove that for acts with finitely many rewards, if $\opm{U}_1\succ^\psi\opm{U}_2$ then $EU_\psi(\opm{U}_1)>EU_\psi(\opm{U}_2)$. For suppose that $\opm{U}_1\succ^\psi\opm{U}_2$. By continuity, if $f$ is the reward function of $\op{U}_1$  (with respect to $\psi$) then it will be possible to find some  act $\opm{V}$ with reward function $g$ such that, for some rewards $r_1$ and $r_2$ with $r_1\succ r_2$:
\begin{itemize}
\item $\opm{V}\succ^\psi\opm{U}$;
\item If $r\neq r_1$ and $r \neq r_2$, $g(r)=f(r)$;
\item $g(r_1)<f(r_1)$; $g(r_2)>f(r_2)$.
\end{itemize} 
This means that we must have $\EU_\psi(\opm{V})\geq \EU_\psi(\opm{U}_2)$; since $\EU_\psi(\opm{V})<\EU_\psi(\opm{U}_1)$, it follows that $EU_\psi(\opm{U}_1)>EU_\psi(\opm{U}_2)$.

This suffices to prove the Born Rule Theorem under the assumption that any act has only finitely many non-null rewards. To extend to the infinite case, let $\opm{U}_1$ and $\opm{U}_2$ be arbitrary acts, and suppose for some $\psi$ that $\opm{U}_1\succ^\psi\opm{U}_2$. By continuity, if $f_1$ and $f_2$ are the reward functions (given $\psi$) of $\opm{U}_1$ and $\opm{U}_2$, it will be possible to find a finite subset $\mc{R}_0$ of $\mc{R}$, and acts $\opm{V}_1,\opm{V}_2$ with reward functions $g_1$, $g_2$, such that:
\begin{itemize}
\item $\opm{V}_1\succ^\psi\opm{V}_2$;
\item $g_i(r)=f_i(r)$ for $r\in \mc{R}_0$;
\item If $r\notin \mc{R}_0$, then $g_1(r)=s$, and $g_2(r)=t$, where $s \succ t$.
\end{itemize}
Since $\opm{V}_1$ and $\opm{V}_2$ have only finitely many non-null rewards, $\EU_\psi(\opm{V}_1)>\EU_\psi(\opm{V}_2)$. But by construction $\EU_\psi(\opm{U}_1)>\EU_\psi(\opm{V}_1)$ and $\EU_\psi(\opm{U}_2)<\EU_\psi(\opm{V}_2)$, so $\EU_\psi(\opm{U}_1)>\EU_\psi(\opm{U}_2)$. $\Box$

\section{Other proposed strategies for action}\label{wallace-otherstrategies}

In the nine years since Deutsch's original paper on decision-theoretic probability, a bewildering variety of alternative strategies for rational action have been proposed in the literature and in discussion. Some of these strategies have independent motivations; some are purely meant as counter-examples; all contradict the Born rule, and so all violate the decision-theoretic axioms of this paper.

This being the case, perhaps there is little need to discuss the alternative strategies: a proof is a proof. On the other hand, it may be instructive to show exactly how some of these alternative proposals violate my axiom scheme: apart from casting light on the motivation for the axioms, this may show how what appear to be coherent and even plausible strategies come apart on close inspection.

The proposed counter-examples, as will become apparent, break into four categories. There are the ``wrong-probability'' rules, which also require an agent to maximise expected utility but with respect to some probability measure other than the Born rule. There are the `no-probability'' rules, which (purportedly) cannot be represented in terms of expected utilities at all. There are what might be called the ``I-don't-want-to-play'' rules, which are not so much positive strategies as arguments against the existence of any strategy. And one special group, the contextual strategies, deserve a category of their own.

\subsection*{Branch counting}
\begin{description}
\item[Description:] each branch is given an equal probability, so that if there are $N$
branches following a particular experiment, each branch is given probability $1/N$. Utility is then maximised with respect to this probability.
\item[Origin:] Has been reinvented innumerable times, but the first proponent may have been Graham, in  \citeN{dewittgraham}. 
\item[Rationale:] Each branch contains a copy of me; none of them can detect, nor care about, their quantum-mechanical weight; so I should not care about that weight either, and so I have no reason to prefer one over another.
\item[Why it is irrational:] The first thing to note about branch counting is that it can't actually be motivated or even defined given the structure of quantum mechanics. There is no such thing as ``branch count'': as I noted earlier, the branching structure emergent from unitary quantum mechanics does not provide us with a well-defined notion of how many branches there are. All quantum mechanics really allows us to say is that there are \emph{some} versions of me for each outcome. 

But within the stylised context of my decision theory, the branch count is defined, so of course (given the representation theorem) the branch counting rule must violate some of my axioms. In fact, it violates the combination of branching indifference and diachronic consistency. For consider two acts $A1$ and $A2$: $A1$ consists of a two-outcome measurement (a spin measurement, say) followed by a reward of utility $r$ in the spin-up branch. $A2$ consists of $A1$, followed by another two-outcome measurement in the spin-up branch. By branching indifference, the agent who gets the reward is indifferent about whether or not he makes a further measurement; by diachronic consistency, then, the original agent is indifferent between $A1$ and $A2$. But the utility of $A1$ (in which there are 2 branches, one of which provides a reward) is $r/2$; the utility of $A2$ is $2r/3$.
\end{description}

\subsection*{The fatness rule}
\begin{description}
\item[Description:] each branch is given a probability proportional to its quantum-mechanical weight multiplied by the mass of the agent in kilograms (such that the total probability is equal to one). Utility is maximised with respect to this probability.
\item[Origin:] David Albert (in conversation, and in his contribution to this volume).
\item[Rationale:] Albert says, tongue-in-cheek, that an agent should care about branches where he is fatter because ``there is more of him'' on that branch. He isn't serious, though: the rule is purely presented as a counter-example.
\item[Why it is irrational:] It violates diachronic consistency. Albert's agent is (ex hypothesi) indifferent to dieting. But he is not indifferent to whether his future selves diet: he wants the ones on branches with good outcomes to gain weight, and the ones on branches with bad outcomes to lose weight. 

This is perhaps a good point to recall the rationale for diachronic consistency: rational action takes place over time and is incompatible with widespread conflict between stages of an agent's life. In the case of the fatness rule, agents have motivation to coerce their future selves --- by hiring ``minders'', say --- into dietary programs that they will resist. Multiply this conflict indefinitely many times (for branching is ubiquitous) and rational action becomes impossible.

(To object ``maybe rational action is impossible in the Everett interpretation'' would, as noted before, be facile. It's perfectly possible for an agent following the Born rule.)
\end{description}

\subsection*{The fake-state rule}
\begin{description}
\item[Description:] The agent maximises expected utilities as for the Born rule, but using a quantum state other than the physically real one.
\item[Origin:] Suggested many times in conversation.
\item[Rationale:] None in particular, though it is often intended to undermine the connection between the ``real'' state and the physics.
\item[Why it is irrational:] It violates state supervenience. There will be cases where two acts produce the same physical state but where one produces a different fake state than the other. (This is inevitable: any two distinct quantum states are invariant under different sets of transformations.) The fake-state rule will then give the acts different utilities; state supervenience rules this out. Or, put another way: the fake state rule assigns different values to the same physical state under two different descriptions.

Note that it is crucial here --- as elsewhere in decision theory --- that the agent has a choice between different actions, and therefore between different sets of histories and weights. Of course, in a deterministic universe it is fixed which action will actually occur, but this does not remove the necessity of defining preferences, and hence indirectly probabilities, over a wide range of actions.
\end{description}
\subsection*{The distributive-justice rule}
\begin{description}
\item[Description:] The agent does not maximise expected utilities at all. He treats his various successors in rather the way that a just ruler would treat his various subjects: in particular, he will not allow the suffering of one even if it brings great advantage to others.
\item[Origin:] Huw Price (this volume).
\item[Rationale:] Any action we choose generates a multitude of individuals; we have a duty to treat them all ethically, and in particular we would not be morally justified in letting one suffer unduly for the others' benefit.
\item[Why it is irrational:] The rule is very underspecified, so it isn't easy to answer this, but on natural precisifications it either violates continuity or is not actually a counterexample to the Born rule.

To expand: a large part of what Price wants can be achieved  by an appropriate utility function. An agent moved by Price's concerns can drastically increase the disutility of bad consequences and scale down the utility of good consequences, with the effect that trade-offs of the sort he considers get a much lower utility and so will tend to be rejected in favour of more equitable options. There is nothing in Everettian decision theory that prevents an agent from making such modifications to their utility function on recognising the ethical consequences of the Everett interpretation.\footnote{Personally, though, I don't feel inclined to. Call me callous.}

If Price wants to hold that \emph{no} amount of suffering, however low-weight the branch on which it occurs, is acceptable, then this strategy will not work, but there is a clash with Continuity. Suppose there are three rewards $r_1$ and $r_2$ with $r_1\succ r_2$, and a (dire) punishment $p$. Price will prefer $r_1$ to $r_2$ but will prefer $r_2$ to $(1-w)r_1+w p$, whatever the value of $w$; clearly this violates continuity.

Now, I think the physical arguments for continuity are pretty unassailable, but it is worth noting that the principle is only really used in my proof precisely to rule out infinite or infinitesimal utilities. (The only other use is for the mathematically convenient but physically tangential purpose of extending the Born rule to the case of infinitely many rewards.) If such utilities are allowed, there is no problem with extending the Born rule to cover even Price's case (though the utility function will have to be modelled in non-standard analysis and the maths will start getting fiddly.) And in fact, precisely the same situation has arisen in \emph{classical} decision theory, and the structure axioms of classical decision theory are selected precisely to rule out the case of infinite (dis)utility. 
\end{description}
\subsection*{The variety rule}
\begin{description}
\item[Description:] An agent prefers $A$ to $B$, but prefers receiving $A$ in half the branches and $B$ in the other half to either $A$ or $B$.
\item[Author:] Suggested in a seminar by Adam Elga in 2004; has not appeared in print as far as I am aware.
\item[Rationale:] An agent may regret having to make one choice or another, and may rather like the idea that one version of himself makes one choice, one another. (In Elga's example, a student prefers physics to history but likes both; that student might prefer to do history in one branch, physics in the other.)
\item[Why it is irrational:] It either violates diachronic consistency, or it isn't a counter-example to the Born rule. 

To expand: suppose you are the agent who chose history. What prevents you changing your mind and switching to physics? It doesn't, after all, hurt your counterpart in the physics branch. This would clearly violate diachronic consistency.

But perhaps you wouldn't choose to switch back. That's to say that although you prefer doing physics to doing history, you prefer doing history \emph{as a result of a situation in which a certain process chose history for you} rather than doing physics \emph{against the result of that process}. In that case, the utility you are assigning to (history-after-process) is higher than the utility you assign to (physics-against-process), and indeed higher than (physics-without-process). The different situations in which you end up doing history count as different rewards.

Exactly analogous situations can arise in classical decision theory. A student might decide that on balance he'd rather do physics than history, but nonetheless resolves to decide by the toss of a coin (because, say, he finds it comforting to have the decision taken from his hands; the reader can probably supply other motivations). That student, again, will place a higher utility on (history after coin toss) than on (physics). 

Of course, if every outcome's utility depended sensitively on the circumstances in which that reward arose, decision theory couldn't get off the ground: there would be no way to define probability without being able to have the same reward available in different acts. But again, this is not specific to quantum decision theory.
\end{description}
\subsection*{The anything-goes rule}
\begin{description}
\item[Description:] Not so much a ``rule'' as a rejection of the need to have one: according to this position, any transitive preference ordering over acts is rationally acceptable.
\item[Origin:] Suggested by Tim Maudlin in seminars on multiple occasions; frequently suggested in conversations.
\item[Rationale:] Everettian quantum theory is deterministic, and we already have a perfectly acceptable deterministic decision theory: its only axiom is transitivity. So any transitive ordering should be fine.
\item[Why it is irrational:] Even in deterministic decision theory, transitivity is not the only constraint. Rational agency is not possible without diachronic consistency; in addition, preference orders have to be defined on actual physical acts, so mathematical modelling of those orders should require an agent to be indifferent between the same state of affairs differently defined. Furthermore, the only interesting decision-theoretic strategies are those which are physically performable in at least an idealised sense. All of the rationality axioms of this paper fit into one of these categories; even in deterministic decision theory, then, they are rationally required.
\end{description}
\subsection*{The curl-up-and-die rule}
\begin{description}
\item[Description:] The converse of the anything-goes rule, this is not so much a ``rule'' for rational action as the claim that \emph{no} rational strategy is possible in Everettian quantum theory.
\item[Origin:] Frequently suggested in conversation.
\item[Rationale:] Various; see below.
\item[Why it is irrational:] Unless there is something concretely wrong with the Born rule, there is no case to be made that no rational strategy is available: the Born rule is available.

I am aware of two general objections to the rationality of the Born rule, though. The first is that it is rationally compulsory for an agent to weight each branch equally; since the Born rule violates this requirement, it cannot be rational (and if only the Born rule is rational, rationality is impossible in an Everettian universe). Arguments are seldom given for the suggestion that this is a rational requirement (I can see that at best it might be a rational \emph{desideratum}, but it's not at all clear to me why, in a universe where it isn't physically possible to obey the requirement, we should be unable to settle for some second-best option). In any case, though (at the risk of repetitiveness) there is no coherent notion of branch count available in quantum mechanics, so it's not even meaningful to talk of ``weighting each branch equally''.

The other objection (frequently made in discussions, and made in print by \citeN{hemmopitowsky}) is that no strategy can be rational if it can be known in advance by those adopting it that some of them (or some of their successors) will make wrong decisions. So in particular, it is a corollary of the Born rule that an agent measuring a long succession of identical quantum systems should regard the observed frequencies as a guide to what state each system is in; but since all sequences of results occur somewhere, some of the agent's successors will get the wrong outcome. 

Now, it is true that some agents will indeed be misled in this way. But there is nothing particularly quantum-mechanical about this. If the universe is spatially infinite (as current observations support), we can guarantee that somewhere in the universe are people as similar to us as you like but whose observed statistics have systematically misled them. Even on Earth, one can fairly easily come up with similar examples. Suppose that the British government declared that it puts some people under (non-covert) surveillance at random, but that there are very few such people: only one in ten million. And suppose it is claimed that the government is lying, and actually puts many more people than that (tens of thousands, say) under surveillance. Then each person in Britain is rational to adopt the strategy: if I am under surveillance, the government is (almost certainly) lying --- even though they know that if the government is not lying, five or six people in Britain will be misled into thinking it was.

Ultimately, some people get unlucky. There is no contradiction between this and the rationality of a decision-theoretic strategy, provided that strategy tells us not to care about the unlucky cases. The Born rule tells us exactly that.
\end{description}

\subsection*{Non-contextual rules}
\begin{description}
\item[Description:] An agent's preferences conform to a probability rule that violates the principle of non-contextuality: that is, it assigns different probabilities to the outcomes of a measurement of operator $\op{X}$ according to whether or not a compatible operator $\op{Y}$ is measured at the same time.
\item[Origin:] Various, but a particularly forceful advocacy can be found in~\cite{hemmopitowsky}.
\item[Rationale:] As is well known, any non-contextual quantum probability rule (and hence, any strategy for rational action expressible in terms of such a rule) can be proved to be the Born rule applied to some (possibly mixed) state.\footnote{This is usually explained in terms of Gleason's Theorem, but this is a rather outdated approach now that POVMs, not PVMs, are widely --- and in my view correctly --- seen as the best way to represent measurements in quantum theory. Most of the mathematical complexity of Gleason's theorem can be dispensed with if we require our probability function to be defined on POVMs and not just PVMs. See \cite{fuchsgleason} for further discussion.} The suspicion, then, is that the decision-theoretic arguments are just a combination of Gleason's theorem (or a relative of it) with an unjustified assumption of non-contextuality.
\item[Why it is irrational:] Probably the easiest way to explain what is wrong with non-contextual rules is that they violate State Supervenience. If we regard measurements as physical processes rather than as primitive, which operator(s) are being measured in a given process is dependent on the interests of the experimenter, and cannot simply be read off from the physics. (Consider the Stern-Gerlach experiment, for instance: is it a measurement of spin, or of position?) For a decision rule to be non-contextual, then, is for a rational agent to prefer a given act to the same act (knowably the same act, in fact) under a different description, which obviously violates State Supervenience (and, I hope, is obviously irrational).

It is fair to note, though, that just as a non-primitive approach to measurement allows one and the same physical process to count as multiple abstractly construed measurements, it also allows one and the same abstractly construed measurement to be performed by multiple physical processes. It is then a non-trivial fact, and in a sense a physical analogue of non-contextuality, that rational agents are indifferent to which particular process realises a given measurement.

In earlier work \cite{decshort,wallaceprobdec} I called this fact \emph{measurement neutrality}. It is indeed a tacit premise in Deutsch's original \citeN{deutschprob} proof of the Born rule, as I argued in \citeN{decshort}. In this paper, it is a theorem (a trivial corollary of the main representation theorem, in fact) that measurement neutrality is rationally required. The short answer as to why is that two acts which correspond to the same abstractly construed measurement can be transformed into the same act via processes to which rational agents are indifference. To see the long answer, re-read sections~\ref{wallace-informalstatement1}--\ref{wallace-formalproof}.

Incidentally, Gleason's theorem (or more accurately its POVM generalisation) is much more directly needed if we wish to generalise the results of this paper to situations where the quantum state is unknown to the agent. The details are somewhat involved; see \citeN{evbook} for an account.
\end{description}

\section{Conclusion}

A rational agent, believing that the Everett interpretation is true and that the quantum state of a given system is $\ket{\psi}$, knows that measurements on that state will generally split his part of the multiverse into multiple branches, with different measurement outcomes, and different versions of the agent, on different branches; he also knows that the relative weights of these branches are given by the Born rule, applied to the post-measurement state of the system and measurement device. Rationality considerations not different in kind to those which apply in single-universe decision making then compel the agent to act as if a set of branches of relative weight $w$ has probability $w$. In other words, he is rationally required to act as if the Born rule is true.

As I noted in the introduction, my focus here is deliberately narrow and I leave it to other chapters in this volume (and to my own work elsewhere) to make the case that such a result suffices to justify the general role of probability in the Everett interpretation. Yet even on its own terms it is a rather remarkable result, as Deutsch's opening quotation notes, and one which to the best of my knowledge has no analogue outside the branching-universe context.

And how does this result actually come about? The decision-theoretic language in which this paper is written is no doubt necessary to make a properly rigorous case and to respond to those who doubt the very coherence of Everettian probability, but in a way the central core of the argument is not decision-theoretic at all. What is really going on is that the quantum state has certain symmetries and the probabilities are being constrained by those symmetries.

This is actually a throwback to an older idea of probability. Quantitative probability has been concerned with symmetry ever since it was applied to the throw of dice in the 17th century: what makes it reasonable to regard each side of a die as equiprobable is that we have no reason to regard one as more probable than another, and what prevents us having reason is the rotational symmetry of the die that maps one side to another. But real dice --- real classical dice, at any rate --- must break the symmetry by their initial conditions, or else how in a deterministic universe could the die land one way rather than another. We then have to impose a certain probability distribution on  the die's initial conditions, and any prospect of a reductive analysis of probability is lost. In Everettian quantum mechanics, there is no one actual outcome, no requirement for the symmetry to be broken by the actual state of the system, and so a program of deriving the probabilities from the symmetries remains viable. The language of decision theory makes rigorous sense of what such a derivation would look like, and shows --- I claim --- that the program can indeed be carried out.

\section*{Acknowledgments}

This work has drawn heavily on conversations and correspondences over a number of years with Harvey Brown, Jeremy Butterfield, David Deutsch, Hilary Greaves, Chris Timpson and Wayne Myrvold, and above all, Simon Saunders.


\begin{thebibliography}{}

\bibitem[\protect\citeauthoryear{Barnum, Caves, Finkelstein, Fuchs, and
  Schack}{Barnum et~al.}{2000}]{barnumetal}
Barnum, H., C.~M. Caves, J.~Finkelstein, C.~A. Fuchs, and R.~Schack (2000).
\newblock {Q}uantum {P}robability from {D}ecision {T}heory?
\newblock {\em Proceedings of the Royal Society of London\/}~{\em A456},
  1175--1182.
\newblock Available online at http://arXiv.org/abs/quant-ph/9907024.

\bibitem[\protect\citeauthoryear{Caves, Fuchs, Manne, and Renes}{Caves
  et~al.}{2004}]{fuchsgleason}
Caves, C.~M., C.~A. Fuchs, K.~Manne, and J.~M. Renes (2004).
\newblock Gleason-type derivations of the quantum probability rule for
  generalized measurements.
\newblock {\em Foundations of Physics\/}~{\em 34}, 193.

\bibitem[\protect\citeauthoryear{Davidson}{Davidson}{1973}]{davidsoninterpreta%
tion}
Davidson, D. (1973).
\newblock Radical interpretation.
\newblock {\em Dialectica\/}~{\em 27}, 313--328.

\bibitem[\protect\citeauthoryear{Davidson}{Davidson}{2004}]{davidsonparadoxes}
Davidson, D. (2004).
\newblock Paradoxes of irrationality.
\newblock In {\em Problems of Rationality}. Oxford: Oxford University Press.

\bibitem[\protect\citeauthoryear{Dennett}{Dennett}{1984}]{dennettelbowroom}
Dennett, D.~C. (1984).
\newblock {\em Elbow Room: the Varieties of Free Will Worth Wanting}.
\newblock Oxford: Oxford University Press.

\bibitem[\protect\citeauthoryear{Dennett}{Dennett}{1987}]{dennettintentional}
Dennett, D.~C. (1987).
\newblock {\em The intentional stance}.
\newblock Cambridge, Mass.: MIT Press.

\bibitem[\protect\citeauthoryear{Deutsch}{Deutsch}{1999}]{deutschprob}
Deutsch, D. (1999).
\newblock Quantum theory of probability and decisions.
\newblock {\em Proceedings of the Royal Society of London\/}~{\em A455},
  3129--3137.
\newblock Available online at http://arxiv.org/abs/quant-ph/9906015.

\bibitem[\protect\citeauthoryear{DeWitt and Graham}{DeWitt and
  Graham}{1973}]{dewittgraham}
DeWitt, B. and N.~Graham (Eds.) (1973).
\newblock {\em The many-worlds interpretation of quantum mechanics}.
\newblock Princeton: Princeton University Press.

\bibitem[\protect\citeauthoryear{Hemmo and Pitowsky}{Hemmo and
  Pitowsky}{2007}]{hemmopitowsky}
Hemmo, M. and I.~Pitowsky (2007).
\newblock Quantum probability and many worlds.
\newblock {\em Studies in the History and Philosophy of Modern Physics\/}~{\em
  38}, 333--350.

\bibitem[\protect\citeauthoryear{Lewis}{Lewis}{1974}]{lewisradical}
Lewis, D. (1974).
\newblock Radical interpretation.
\newblock {\em Synthese\/}~{\em 23}, 331--44.
\newblock Reprinted in David Lewis, \textit{Philosophical Papers}, Volume I
  (Oxford University Press, Oxford, 1983).

\bibitem[\protect\citeauthoryear{Lewis}{Lewis}{1980}]{lewischance}
Lewis, D. (1980).
\newblock A subjectivist's guide to objective chance.
\newblock In R.~C. Jeffrey (Ed.), {\em Studies in Inductive Logic and
  Probability}, Volume~II. Berkeley: University of California Press.
\newblock Reprinted in David Lewis, \textit{Philosophical Papers}, Volume II
  (Oxford University Press, Oxford, 1986).

\bibitem[\protect\citeauthoryear{Lewis}{Lewis}{2005}]{lewisondeutsch}
Lewis, P.~J. (2005).
\newblock Probability in {E}verettian quantum mechanics.
\newblock Available online at http://phil-sci.pitt.edu.

\bibitem[\protect\citeauthoryear{Saunders}{Saunders}{1998}]{saundersprobabilit%
y}
Saunders, S. (1998).
\newblock {T}ime, {Q}uantum {M}echanics, and {P}robability.
\newblock {\em Synthese\/}~{\em 114}, 373--404.

\bibitem[\protect\citeauthoryear{Saunders and Wallace}{Saunders and
  Wallace}{2008}]{saunderswallace}
Saunders, S. and D.~Wallace (2008).
\newblock Branching and uncertainty.
\newblock {\em British Journal for the Philosophy of Science\/}~{\em 59},
  293--305.

\bibitem[\protect\citeauthoryear{Savage}{Savage}{1972}]{savage}
Savage, L.~J. (1972).
\newblock {\em The foundations of statistics\/} (2nd ed.).
\newblock New York: Dover.

\bibitem[\protect\citeauthoryear{Wallace}{Wallace}{2001}]{wallacestatmech}
Wallace, D. (2001).
\newblock {I}mplications of {Q}uantum {T}heory in the {F}oundations of
  {S}tatistical {M}echanics.
\newblock Available online from http://philsci-archive.pitt.edu.

\bibitem[\protect\citeauthoryear{Wallace}{Wallace}{2003}]{decshort}
Wallace, D. (2003).
\newblock {E}verettian rationality: defending {D}eutsch's approach to
  probability in the {E}verett interpretation.
\newblock {\em Studies in the History and Philosophy of Modern Physics\/}~{\em
  34}, 415--439.
\newblock Available online at http://arxiv.org/abs/quant-ph/0303050 or from
  http://philsci-archive.pitt.edu.

\bibitem[\protect\citeauthoryear{Wallace}{Wallace}{2005}]{wallacebranching}
Wallace, D. (2005).
\newblock Language use in a branching universe.
\newblock Forthcoming; Available online from http://philsci-archive.pitt.edu.

\bibitem[\protect\citeauthoryear{Wallace}{Wallace}{2006}]{wallaceepist}
Wallace, D. (2006).
\newblock Epistemology quantized: circumstances in which we should come to
  believe in the {E}verett interpretation.
\newblock Forthcoming in \emph{British Journal for the Philosophy of Science}.
  Available online from http://philsci-archive.pitt.edu.

\bibitem[\protect\citeauthoryear{Wallace}{Wallace}{2007}]{wallaceprobdec}
Wallace, D. (2007).
\newblock Quantum probability from subjective likelihood: Improving on
  {D}eutsch's proof of the probability rule.
\newblock {\em Studies in the History and Philosophy of Modern Physics\/}~{\em
  38}, 311--332.

\bibitem[\protect\citeauthoryear{Wallace}{Wallace}{2010}]{evbook}
Wallace, D. (2010).
\newblock {\em The Everett Interpretation}.
\newblock Oxford University Press.

\bibitem[\protect\citeauthoryear{Zurek and Paz}{Zurek and Paz}{1994}]{zurek94}
Zurek, W.~H. and J.~P. Paz (1994).
\newblock Decoherence, chaos and the second law.
\newblock {\em Physical Review Letters\/}~{\em 72\/}(16), 2508--2511.

\end{thebibliography}
\end{document}